\def\gr{$\gamma$-ray }
\def\grs{$\gamma$-rays }
\def\be{\begin{equation}}
\def\ee{\end{equation}}

\documentclass{aa}
\usepackage{times}
\usepackage{graphics}
\begin{document}
\thesaurus{03(02.09.1; 02.18.5; 02.20.1; 11.02.1; 13.07.3)}

\title{On the conversion of blast wave energy into radiation in active 
galactic nuclei and gamma-ray bursts}
\author{Martin Pohl and Reinhard Schlickeiser}
\institute{Institut f\"ur Theoretische Physik,
Lehrstuhl IV: Weltraum- und Astrophysik, Ruhr-Universit\"at Bochum,
D-44780 Bochum, Germany}
\date{Received ; accepted }
\offprints{mkp@tp4.ruhr-uni-bochum.de}
\titlerunning{Channelled blast wave model for AGN}
\maketitle

\begin{abstract}

It has been suggested that relativistic blast waves may power the 
jets of AGN and gamma-ray bursts (GRB).
We address the important issue how the kinetic energy of 
collimated blast waves is converted into radiation. It is shown that
swept-up ambient matter is quickly isotropised in the blast wave frame
by a relativistic two-stream instability, which provides relativistic
particles in the jet without invoking any acceleration process. The fate
of the blast wave and the spectral evolution of the emission of the 
energetic particles is therefore solely determined by the initial 
conditions.
We compare our model with existing multiwavelength data of AGN and
find remarkable agreement.


\end{abstract}

\section{Introduction}

Shortly after the detection of more than 60 blazar-type active galaxies
as emitters of MeV--GeV \grs by EGRET on board the
Compton Gamma-Ray Observatory (CGRO) (von Montigny et al. \cite{cvm95},
Mukherjee et al. \cite{muk97}), observations with ground-based
\v Cerenkov telescopes have shown that at least for BL Lacertae objects
the \gr spectrum can be traced to more than a TeV observed photon energy
(Punch et al. \cite{pun92}, Quinn et al. \cite{qui96}, Catanese et al.
\cite{cat98}, Krennrich et al. \cite{kre97}). In these sources the bulk of luminosity is often emitted in the form of $\gamma$-rays.
The typical \gr spectra of blazars are well described
by power laws in the energy range between a few MeV and a few GeV
(Mukherjee et al. \cite{muk97}), but 
display a break around an MeV (McNaron-Brown et al. \cite{mnb95}).
At GeV energies the spectra show a tendency of getting harder during
outbursts (Pohl et al. \cite{poh97}). The \gr emission of the typical
blazar is highly variable on all timescales down to the observational
limits of days at GeV energies and hours at TeV energies
(e.g. Mattox et al. \cite{mat97}, Gaidos et al. \cite{gai96}).
Large amplitude variability at TeV energies is generally not associated with
corresponding variability at GeV energies (Lin et al. \cite{lin94},
Buckley et al. \cite{buc96}).

Correlated monitoring of the sources at lower frequencies 
has demonstrated that \gr outbursts are accompanied by activity
in the optical and radio band (Reich et al. \cite{rei93},
M\"ucke et al. \cite{mue96}, Wagner \cite{wag96}). Emission constraints
such as the compactness limit and the Elliot-Shapiro relation (Elliot \&
Shapiro \cite{es74}) are violated in some \gr blazars, even when these
limits are calculated in the Klein-Nishina limit (Pohl et al. \cite{poh95},
Dermer \& Gehrels \cite{dg95}), which implies relativistic bulk motion
within the sources. This conclusion is further supported by 
observations of apparent superluminal motion in many \gr blazars
(e.g. Pohl et al. \cite{poh95}, Barthel et al. \cite{bar95},
Piner \& Kingham \cite{pk97a,pk97b}). 

Published models for the \gr emission of blazars are usually based
on inverse Compton scattering of soft target photons by highly relativistic
electrons in the jets of these sources. The target photons may come directly
from an accretion disk (Dermer et al. \cite{dsm92},
Dermer \& Schlickeiser \cite{ds93}), or may be rescattered accretions disk
emission (Sikora et al. \cite{sbr94}), or may be produced in the
jet itself via synchrotron radiation (e.g. Bloom \& Marscher \cite{bm93} and 
references therein). In all the above models the jet and its environment
is assumed to be optically thin for \grs in the MeV--GeV range. A
common property of the inverse Compton models is the emphasis on the
radiation process and the temporal evolution of the electron spectrum
and the neglect of the problem of electron injection and acceleration
(for an exception see Blandford \& Levinson \cite{bl95}).
The modeling of the multifrequency spectra of blazars often requires
a low energy cut-off in the electron injection spectrum, but no high
energy cut-off.
The usual shock or stochastic electron acceleration processes would have to be
very fast to compete efficiently with the radiative losses at
high electron energies, and a high energy cut-off should occur
under realistic conditions, but no low energy cut-off (e.g.
Schlickeiser \cite{sch84}), in contrast to the requirements of the 
spectral modeling.

When efficient, but possibly slower, acceleration of protons is assumed,
photomeson production on ambient target photons can provide many
secondary electrons and positrons with a spectrum as required by the
multifrequency modeling (Kazanas \& Ellison \cite{ke86}, Sikora et al.
\cite{sik87}, Mannheim \& Biermann \cite{mb92}). Such systems are usually
optically thick and a pair cascade develops. In these models multi-TeV
\gr emission can be easily produced, but the observed short
variability timescale places extreme constraints on the magnetic
field strength, because the proton gyroradius has to be much smaller than
the system itself, and on the Doppler factor, because the intrinsic
timescale for switching off the cascade is linked to the observed
soft photon flux via the energy loss rate for photomeson production.
For the most rapid \gr outburst of Mrk 421 (Gaidos et al. \cite{gai96})
this leads to $B\gg 10\ {\rm G}$ and $D\gg 100$.

Another class of models features hadronic collisions of a collimated
proton jet with BLR clouds entering the jet (e.g. Dar \& Laor \cite{dl97};
Beall \& Bednarek \cite{bb99}).
These models have two theoretical difficulties: the proton beam is weak
compared with the background plasma and therefore quickly stopped by a
two-stream instability. Also the BLR clouds are usually optically thick
and thus the efficiency of the system is drastically reduced.

In this paper we consider a strong electron-proton beam that sweeps
up ambient matter and thus becomes energized. The basic scenario is similar
to the
blast wave model for $\gamma $-ray bursts (GRBs) which successfuly explains
the time dependence of the X-ray, optical and radio afterglows (e.g. 
Wijers et al. \cite{wmr97}, Vietri \cite{vie97a},
Waxman \cite{wax97a}). There the
apparent release of $\simeq 10^{52}E_{52}$ erg of energy in a
small volume leads to the formation
of a relativistically expanding pair fireball that transforms most of 
the explosion energy into kinetic energy of baryons in a relativistic
blast wave. We assume that a similar generation
process also powers the relativistic outflows in active galactic
nuclei (AGN), but that the outflow is not spherically symmetric
and highly channelled along magnetic flux 
tubes into a small fraction of the full solid angle due
to the structure of the medium surrounding the point of energy release.

We shall address the important issue how the kinetic
energy of such channelled relativistic blast waves is converted into radiation.
Existing radiation modelling of GRBs and AGNs (see e.g. Vietri \cite{vie97b};
Waxman \cite{wax97b}; B\"ottcher \& Dermer \cite{bd98}) are very unspecific on 
this crucial point. Typically it is assumed that a fraction $\xi $ of the 
energy in nonthermal baryons in the blast wave region at position $x$ 
is transformed into 
a power-law distribution
of ultrarelativistic cosmic ray protons with Lorentz factors
$\Gamma (x)\le \gamma _{\rm CR}\le \gamma _{\rm max}$ in the fluid frame
comoving with a small element of the blast wave region that travels with
the bulk Lorentz factor $\Gamma (x)=\Gamma _0(x/x_0)^{-g}$ after the
deceleration radius $x_0=2.6\cdot 10^{16}(E_{52}/n_0\Gamma _{0,300}^2)^{1/3}$
cm through the surrounding medium of density $n_0$.
For electrons it is argued (Katz \& Piran \cite{kp97}, Panaitescu \&
M\'esz\'aros \cite{pm98}) that their minimum Lorentz factor is
$\gamma _{\rm e,min}=(m_{\rm p}/m_{\rm e})\Gamma $
since they are in energy equilibrium with the protons.
Here we investigate this transfer mechanism from the channelled blast wave
to relativistic protons and electrons in more detail. In Sect. 2
we consider the penetration of a blast wave consisting of 
cold protons and electrons with density $n_{\rm b}$ with the surrounding "interstellar"
medium consisting also of cold protons and electrons on the basis of 
a two-stream instability. Viewed from the coordinate system comoving
with the blast wave, the interstellar protons and electrons represent
a proton-electron beam propagating with the 
relativistic speed $V(x)=-c(1-\Gamma ^{-2}(x))^{1/2}$
antiparallel to the $x$-axis. We examine the stability of this beam
assuming that the background magnetic field is uniform and directed along
the x-axis. We demonstrate that very quickly the beam excites low-frequency
magnetohydrodynamic plasma waves, mainly
Alfv\'en-ion-cyclotron and Alfv\'en-Whistler waves. These plasma waves 
quasi-linearly isotropise
the incoming interstellar protons and electrons in the blast wave plasma. 
In Sect. 3 we investigate the interaction processes of these isotropised
protons and electrons (hereafter referred to as primary protons and electrons),
which 
have the relativistic Lorentz factor $\Gamma (x)$, with the blast wave
protons and electrons. Since the primary protons carry much more 
momentum than the primary electrons, inelastic collisions between
primary protons and the blast wave protons generate neutral and charged
pions which decay into gamma rays, secondary electrons, positrons and 
neutrinos. Both, the radiation products from these interactions, and the 
resulting cooling of the primary particles in the blast wave plasma, are 
calculated. By transforming to the laboratory frame we calculate the time
evolution of
the emitted multiwavelength spectrum for an outside observer under different 
viewing angles. Momentum conservation leads to a deceleration of the blast
wave that is taken into account self-consistently. Since we do not consider 
any re-acceleration of particles in the blast wave, the evolution of
particles and the blast wave is completely determined by the initial
conditions. 

\section{Two-stream instability of a proton-electron beam}
\subsection{Physical model and basic equations}

As sketched in Fig. 1 we consider in the laboratory frame (all physical
quantities in this system are indexed with $*$)
the cold blast wave electron-proton
plasma of density $n^*_{\rm b}$ and thickness $d^*$ in $x$-direction running into 
the cold interstellar medium of density $n^*_0$, consisting also of 
electrons and protons, parallel to the uniform magnetic field of
strength $B$. In the comoving frame 
the total phase space distribution
function of the plasma in the blast wave region at the start thus is
\begin{eqnarray}
f^*(\vec{p}^*,t=0)=&\,&{1\over 2\pi p_{\perp }^*}n^*_0\delta \bigl(
p^*_{\perp }\bigr)\delta \bigl(p^*_{\parallel }\bigr) \nonumber \\
&+& 
{1\over 2\pi p_{\perp }^*}n^*_{\rm b}\delta \bigl (p^*_{\perp }\bigr)
\delta \bigl(p^*_{\parallel }-P\bigr),
\label{1}\end{eqnarray}
where 
$$P=\Gamma mV=\Gamma m\beta c=mc \sqrt{\Gamma ^2-1)}.$$
The Lorentz transformations (e.g. Hagedorn \cite{hag73}, p.41) to the blast wave
rest frame with momentum and energy variables $(p_{\perp }, p_{\parallel },
E)$ are
\be
p^*_{\perp }=p_{\perp },\;\;\;
p^*_{\parallel }=\Gamma (p_{\parallel }+\, \beta {E\over c}),\;\;\;
E^*=\Gamma (E+\, \beta cp_{\parallel })\ .\ee

\begin{figure}
\resizebox{\hsize}{!}{\includegraphics{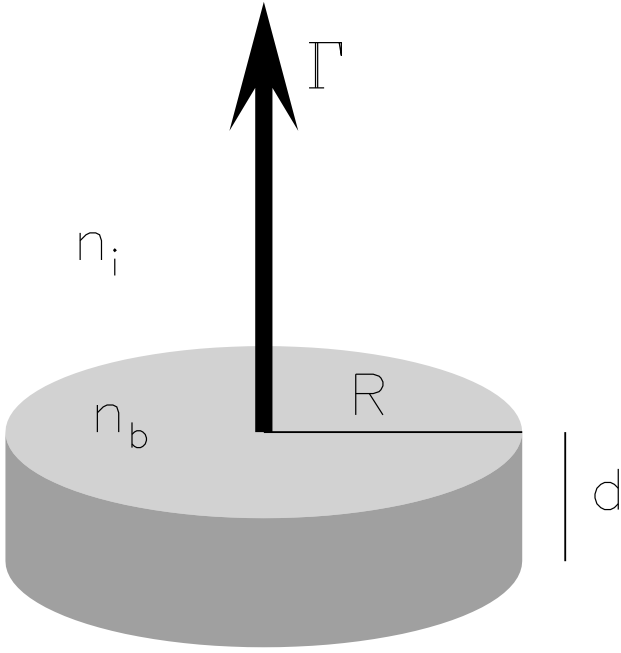}}
\caption{Sketch of the basic geometry. The thickness of the
channeled blast wave $d$, measured in its rest frame, is much smaller than its halfdiameter. The blast wave
moves with a bulk Lorentz factor $\Gamma$ through ambient matter
of density $n_{\rm i}$.}
\label{sketch}
\end{figure}

Using the invariance of the phase space densities we obtain for
the phase space density in the blast wave rest frame
\begin{eqnarray}
 f(\vec{p},t=0)&=&f^*(\vec{p}^*(\vec{p}),t=0) \nonumber \\ &=&
{{n_{\rm i}\delta \bigl(
p_{\perp }\bigr)\delta \bigl(p_{\parallel }-P\bigr)}\over {2\pi p_{\perp }}}+\; 
 {{n_{\rm b}\delta \bigl (p_{\perp }\bigr)
\delta \bigl(p_{\parallel }\bigr)}\over {2\pi p_{\perp }}}
\ ,\label{dist_f}\end{eqnarray}
where the number densities are
$$n_{\rm i}=\Gamma n_0^*,\;\;\, n_{\rm b}=n_{\rm b}^*/\Gamma \ .$$
The blast wave density transformation is inverse to that for the density
of interstellar matter because of the different rest systems.

In terms of
spherical momentum coordinates $p_{\parallel }=p\mu $
and $p_{\perp }=p\sqrt{1-\mu ^2}$ the distribution function (\ref{dist_f}) reads
\begin{eqnarray}
f({p},t=0)&=&
{1\over 2\pi p^2}n_{\rm i}\delta \bigl(\mu +1\bigr)\delta \bigl(p-P\bigr) +\; 
{1\over 4\pi p^2}n_{\rm b}\delta \bigl (p\bigr) \nonumber \\
&=&n_{\rm b}f_{\rm b}(p,\mu ,t=0)+\, n_{\rm i}f_i(p,\mu ,t)
\ ,\label{dist_f1}
\end{eqnarray}
where $\mu =p_{\parallel }/p$ is
the cosine of the pitch-angle of the particles in the magnetic field
$B$ pointing to positive $z$-direction. Obviously, the incoming
interstellar protons and electrons are a beam propagating antiparallel
to the magnetic field direction in the blast wave plasma.
For ease of exposition we assume that the blast wave plasma is
cold, i.e. $f_{\rm b}\propto \delta (p)$. 

The blast wave density $n^*_{\rm b}=10^8n^*_{\rm b,8}$ is much larger than 
the interstellar gas density $n^*_0=1n^*_{\rm i}$.
We want to calculate the time $t_{\rm R}$ it takes to isotropise the incoming
interstellar protons and electrons; if this relaxation time is much
smaller than $d/c$ an isotropic distribution of primary protons and
electrons is effectively generated.

If the beam distribution function $f_i$ in Eq. (\ref{dist_f1}) is 
unstable in the blast wave plasma it will excite plasma waves.
Therefore we have to find the dispersion relation $\omega =\omega (k)$
which are perturbations of the form
\be
\delta \vec{E}=\delta \vec{E}_0\, \exp (\imath (\vec{k}\cdot \vec{x}-\omega t))
\ .
\ee
It has been shown (Kennel \& Wong \cite{kw67}, Tademaru \cite{tad69})
that waves
propagating obliquely to the magnetic field are thermally 
damped by the background (=blast wave) plasma, so that we concentrate 
on parallel ($\vec {k}||\vec {B}$) waves. One finds two implicit
dispersion relations (Baldwin et al. \cite{bbw69};
Achatz et al. \cite{asl90}),
one for electrostatic waves $(\delta \vec {E}|| \vec {B})$
\be
0=\omega ^2-\; \omega_{\rm p,p}^2-\; \omega_{\rm p,e}^2
-\; {n_{\rm i}\over n_{\rm b}\Gamma ^3}\left(1+{{m_{\rm e}}\over {m_{\rm p}}}\right)
{\omega_{\rm p,e}^2\omega ^2\over [\omega +kV]^2}\ ,
\ee
and one for electromagnetic $(\delta \vec {E} \perp \vec {B})$

\begin{eqnarray}
&{n_{\rm i}\omega_{\rm p,e}^2[\omega +kV]\over n_{\rm b}\Gamma } \,
\bigl[[\omega +kV-{{\Omega_{\rm e}}\over {\Gamma}}]^{-1}+
\;{{m_{\rm e}}\over {m_{\rm p}}}[\omega +kV-{{\Omega_{\rm p}}\over {\Gamma}}]^{-1}\bigr]\hfil 
\nonumber \\
&+\left[ c^2k^2-\; \omega ^2+\; {\omega \omega_{\rm p,p}^2\over \omega -\Omega_{\rm p}}
+\; {\omega \omega_{\rm p,e}^2\over \omega -\Omega_{\rm e}}\right]
=0\ ,\hfill \label{disp}\end{eqnarray}
where $\omega_{\rm p,j}=\sqrt{4\pi n_{\rm b}q_{\rm j}^2/m_{\rm j}}$ denotes the plasma frequency
and $\Omega _{\rm j}=q_{\rm j}B/m_{\rm j}c$ the nonrelativistic gyrofrequency of
particle species $j$. In our notation negative (positive ) frequencies
$\omega_{\rm R}$ refer to right-handed (left-handed) circularly polarised
waves, respectively. Moreover, positive phase speeds $(\omega_{\rm R}/k>0)$
indicate forward propagating waves and negative phase speeds
$(\omega_{\rm R}/k<0)$ backward propagating waves.
Here we will concentrate on the excitation of the electromagnetic
waves. The excitation of electrostatic turbulence and its backreaction
on the particle distribution function will be discussed in a forthcoming 
publication.
Because
\be
n_{\rm i}/n_{\rm b}=\Gamma ^2n^*_0/n^*_{\rm b}=
10^{-4}\Gamma _{100}^2n_{\rm i}^*/n^*_{\rm b,8}<<1\ ,\ee
the beam is weak
and the contributions from the beam in Eq.(\ref{disp}) can
be described as perturbations
to the dispersion relation in a single component electron-proton plasma
($n_{\rm i}=0$). Its solutions 
\be
c^2k^2=J(\omega_{\rm R})\equiv 
\omega_{\rm R}^2\Bigl(1+\; {\omega_{\rm p,e}^2\over \Omega_{\rm e}^2}
+\; {\omega_{\rm p,e}^2\over (\omega_{\rm R} -\Omega_{\rm p})\Omega_{\rm e}}\Bigr)\ \ee
are 

\noindent (a) Alfv\'en waves
\be
\omega_R^2=V_{\rm A}^2k^2\ \label{alfven}\ee
  at 
frequencies $|\omega_{\rm R}|<<\Omega_{\rm p}$, where $V_{\rm A}=B(4\pi m_{\rm p}n_{\rm b})^{-1/2}$
is the Alfv\'en speed and $\omega_{\rm R}=\Re (\omega )$. The dispersion relation 
(\ref{alfven})
accounts for four types of Alfv\'en waves: forward and backward moving, right-handed
and left-handed polarised;

\noindent (b)
Whistler waves 
\be
\omega_{\rm R}=\Omega_{\rm e}k^2c^2/\omega_{\rm p,e}^2\ \label{whist}\ee 
at frequencies
between $\Omega_{\rm e}<\omega_{\rm R}<-\Omega_{\rm p}$. The dispersion relation (\ref{whist})
describes right-handed polarised waves (because $\Omega_{\rm e}<0$) that
propagate forward for negative $k<0$ and backward for positive $k>0$.

All of these are stable 
($\psi =\Im (\omega )=0$) if the beam particles are absent. As is explained
in Achatz et al.
(\cite{asl90}) the beam protons and electrons, which under the 
given weak-beam condition do not affect each other, are 
however able to trigger each its instability which occurs if the resonance
conditions
\be
\omega_{\rm R}=-kV +\; (\Omega_{\rm p}/\Gamma ),\;\;\;
\omega_{\rm R}=-kV +\; (\Omega_{\rm e}/\Gamma ),\;\;\; \ee
can be satisfied.

The time-dependent behaviour of the intensities $I(k,t)$ 
of the excited waves is given by (Lerche \cite{ler67}, Lee \& Ip \cite{li87})
\be
{\partial I_{\rm n}\over \partial t}=\psi _{\rm n} I_{\rm n} \ ,
\label{15}\ee
where the growth rate $\psi $ is
\begin{eqnarray}
& &\psi (k)\simeq\pi^2c^3[{\partial J(\omega_{\rm R})\over \partial \omega_{\rm R}}]^{-1}
\rm {sgn} (k)\sum _i\omega _{p,i}^2(m_ic)^3 \ \times\nonumber \\
& &\int_{E_i}^\infty dE\ 
{{E^2-1-({E\over N}-x_i)^2}\over \sqrt{E^2-1}}
{\partial f_i\over \partial \mu }
\delta (\mu +{{x_i}\over \sqrt{E^2-1}})\ ,\label{psi}\end{eqnarray}
where $N=ck/\omega_{\rm R}$ is the index of refraction and further
$E_i=\sqrt{1+x_i^2}$ whith
$x_i=\Omega _{\rm i,0}/kc$.

To describe the influence of these excited waves 
on the beam particles we use the quasilinear Fokker-Planck
equation (e.g. Schlickeiser \cite{sch89}) for the resonant wave-particle 
interaction. For Alfv\'en waves and for Whistler waves the index of refraction
\be
|N_{\rm A}|={\sqrt{J(\omega_{\rm R})}\over |\omega_{\rm R}|}\simeq
1-{\omega_{\rm p,e}^2\over \Omega_{\rm e}\Omega_{\rm p}}\simeq (c/V_{\rm A})^2>>1\ee
\be
|N_{\rm W}|\simeq ({\omega_{\rm p,e}\over \Omega_{\rm e}})^2>>1\ee
is large compared to unity, so that the Lorentz force associated
with the magnetic field of the waves is much larger than the force
associated with the electric field, so that on the shortest time scale
these waves scatter the particles in pitch angle $\mu $ but conserve their
energy, i.e. they isotropise the beam particles. The Fokker-Planck equation
for the phase space density then reads
\be
{\partial f_i\over \partial t}=
{\partial \over \partial \mu }\bigl [
D_{\rm \mu\mu }{\partial f_i\over \partial \mu }\bigr]\ , \label{17}\ee
where the pitch angle Fokker-Planck coefficient is determined by 
the wave intensities $I_{\rm n}$
$$D_{\rm \mu\mu }\simeq 
\sum _n {{\pi\,\Omega _i^2\,(1-\mu ^2)}\over
{2\,B^2}}
\int_{-\infty}^{\infty} dk\ I_{\rm n}(k)\,\delta 
(\omega_{\rm R}-kv\mu -\Omega _i)$$
\be\, \label{Dmumu}\ee
 Using the dispersion relation of Alfv\'en waves ($\omega_{\rm R}\simeq V_{\rm A}k$),
Eq. (\ref{Dmumu}) indicates that beam protons and electrons  
resonate with waves at wavenumbers given by the inverse of
their Larmor radii times $\mu $, 
$k\simeq \Omega _i/\Gamma v\mu =(\mu R_i)^{-1}$. 
For protons these wavenumbers correspond to Alfv\'en waves.  For electrons
these wavenumbers correspond to Alfv\'en waves if the bulk Lorentz factor
is above $|\mu |(m_{\rm p}/m_{\rm e})=1836|\mu |$, and 
Whistler waves for smaller Lorentz factors.
We concentrate here on the isotropisation by Alfv\'en waves mainly for
two reasons:

\noindent (1) the bulk of the momentum of the inflowing interstellar
particles is carried by the protons so that they are energetically more
important than the electrons;

\noindent (2) for Lorentz factors $\Gamma >|\mu | m_{\rm p}/m_{\rm e}$ the isotropisation
of electrons is also caused by scattering with Alfv\'en waves;
for smaller
Lorentzfactors the mistake one makes in representing the Whist\-ler dispersion
relation still by the Alfv\'en dispersion relation is relatively small.

\subsection{Self excited Alfv\'en waves}

For Alfv\'en waves (\ref{alfven}) we obtain 
\be
{\partial J(\omega_{\rm R})\over \partial \omega_{\rm R}}\simeq 
2\omega_{\rm R}c^2/V_{\rm A}^2= 2c^2k/V_{\rm A}\ee
which is positive for forward ($k>0$) moving waves and negative
for backward ($k<0$) moving waves. According to Eq. (\ref{psi}) we obtain
for the growth rate of forward (+) and backward (-) moving Alfv\'en waves
\be
\psi _{\rm \pm }=\pm \psi \label{20a}\ee
with 
\begin{eqnarray}
\psi =&  {\pi^2V_{\rm A}\over c^2|k|}\sum_i \omega _{p,i}^2
(m_ic)^3\ \times\ \nonumber \\
&\int_{E_i}^\infty dE {E^2-E_i^2\over \sqrt{E^2-1}}
{\partial f_i\over \partial \mu }\delta (\mu + x_i(E^2-1)^{-1/2})\ 
\label{alfpsi}\end{eqnarray}
Now it is convenient to introduce the 
normalised phase space distribution function of the
beam particles
\be
f_i(p,\mu ,t)={\delta (E-\Gamma )\over 
2\pi (m_ic)^3\Gamma (\Gamma ^2-1)^{1/2}}
F_i(\mu ,t)\ , \label{21}\ee
where $E=\sqrt {1+(p/mc)^2}$,
so that Eq. (\ref{alfpsi}) reduces to 
\begin{eqnarray}
\psi =  &{\pi \over 2}{V_{\rm A}\over c^2|k|}\sum _i{\omega _{\rm p,i}^2\over \Gamma }
H[\Gamma -\sqrt{1+x_i^2}][1-{x_i^2\over \Gamma^2-1}]\ \times\ \nonumber \\
 &{\partial F_i\over \partial \mu }\delta (\mu -x_i(\Gamma ^2-1)^{-1/2})
\ \label{22}\end{eqnarray}
where $H[x]=1(0)$ for $x>(<)0$ is the step function. Summing over 
protons and electrons separately, and introducing
the proton and electron Larmor radii, $R_{\rm p}= c\sqrt{\Gamma ^2-1}/\Omega_{\rm p}$,
$R_{\rm e}=c\sqrt{\Gamma ^2-1}/|\Omega_{\rm e}|$, respectively, 
Eq. (\ref{22}) becomes
\begin{eqnarray} & \psi = b(|k|) \sum_{\rm x=e,p}\ \nonumber \\
 & {m_{\rm e}\over m_x}[1-(kR_{\rm x})^{-2}]H[|k|-R_{\rm x}^{-1}]
{\partial F_{\rm x}\over \partial \mu }\delta (\mu +(R_{\rm x}k)^{-1})\qquad \end{eqnarray}
with
\be
b(|k|)\equiv {\pi \over 2}{V_{\rm A}\over c}{n_{\rm i}\over n_{\rm b}\Gamma }
{\omega_{\rm p,e}^2\over c|k|}\ .\ee
According to Eqs. (\ref{15}) and (\ref{20a}) written in the form
\be
{\partial I_+(t)\over \partial t}=\psi I_+(t)\ \ee
\be
{\partial I_-(t)\over \partial t}=-\psi I_-(t)\ \ee
we derive
\be
{\partial (I_+(t)I_-(t))\over \partial t}=0\ \ee
and 
\be
{\partial [I_+(t)-I_-(t)]\over \partial t}=\psi [I_+(t)+I_-(t)]\ \ee
or
when integrated,
\be
I_+(t)I_-(t)=I_+(t=0)I_-(t=0)\ \label{27a}\ee
and 
\be
[I_+(t)-I_-(t)]-[I_+(t=0)-I_-(t=0)]=Z(k)\ \label{27b}\ee
with
\begin{eqnarray}
& &Z(k)\equiv 
b(|k|) \sum_{\rm x=e,p} {m_{\rm e}\over m_{\rm x}}[1-(kR_{\rm x})^{-2}]H[|k|-R_{\rm x}^{-1}]\ \times
\nonumber \\
& &\int_0^tdt^{'}\,{\partial F_{\rm x}\over \partial \mu }\delta (\mu +(R_{\rm x}k)^{-1})\left(I_+(t^{'})+I_-(t^{'})\right)\quad\label{27c}\end{eqnarray}
which relates the wave intensities of forward (+) and backward (-)
moving waves at any time $t$ to the initial wave intensities at time $t=0$.

In terms of the normalised distribution functions (\ref{21}) the proton and
electron distribution functions evolve according to Eqs.
(\ref{17})-(\ref{Dmumu}),
\be
{\partial F_{\rm p}\over \partial t}=
{\partial \over \partial \mu }\bigl[a_{\rm p}(|\mu |)\sum _{\pm }
I_{\pm }(|R_{\rm p}\mu |^{-1}){\partial F_{\rm p}\over \partial \mu }\bigr] \ \label{28}\ee
\be
{\partial F_{\rm e}\over \partial t}=
{\partial \over \partial \mu }\bigl[a_{\rm e}(|\mu |)\sum _{\pm }
I_{\pm }(|R_{\rm e}\mu |^{-1}){\partial F_{\rm e}\over \partial \mu }\bigr] \ \label{29}\ee
with 
\be
a_{\rm p}(|\mu |)\equiv {\pi \over 2} {v(1-\mu ^2)\over B^2R_{\rm p}^2|\mu |},\;\;\;
a_{\rm e}(|\mu |)\equiv {\pi \over 2}{v(1-\mu ^2)\over B^2R_{\rm e}^2|\mu |}\ \ee
Integrating Eqs. (\ref{28}) and (\ref{29}) over $t$ and $\mu $ gives because
of $a_{\rm p,e}(\mu =-1)=0$
\begin{eqnarray}
\int _{-1}^{\mu }d\mu ^{'}[F_{\rm p,e}(\mu ^{'},t)-F_{\rm p,e}(\mu ^{'},t=0)]& \qquad\nonumber \\
=a_{\rm p,e}(|\mu |)\sum _{\pm }\int_0^tdt^{'}I_{\pm }(|R_{\rm p,e}\mu |^{-1},t^{'})
{\partial F_{\rm p,e}\over \partial \mu }\ & \quad \label{31}\end{eqnarray}
respectively.
Evaluating Eq. (\ref{31}) at $\mu =-(R_{\rm p}k)^{-1}$ for protons and at
$\mu =(R_{\rm e}k)^{-1}$ we obtain
\begin{eqnarray}
\sum _{\pm }\int_0^tdt^{'}I_{\pm }(|k|,t^{'})
{\partial F_{\rm p}\over \partial \mu }\delta (\mu +(R_{\rm p}k)^{-1})=& \quad \nonumber \\
a_{\rm p}^{-1}(|R_{\rm p}k|^{-1})
\int _{-1}^{-1/(R_{\rm p}k)}d\mu [F_{\rm p}(\mu ,t)-F_{\rm p}(\mu ,t=0)]\ & \ \label{32}\end{eqnarray}
and
\begin{eqnarray}
\sum _{\pm }\int_0^tdt^{'}I_{\pm }(|k|,t^{'}){\partial F_{\rm e}\over \partial \mu }
\delta (\mu -(R_{\rm e}k)^{-1})=& \qquad \nonumber \\
a_{\rm e}^{-1}(|R_{\rm e}k|^{-1})
\int_{-1}^{1/R_{\rm e}k}d\mu [F_{\rm e}(\mu ,t)-F_{\rm e}(\mu ,t=0)]\ & \quad \label{33}\end{eqnarray}
Eqs. (\ref{32}) and (\ref{33})  can be
inserted in Eq. (\ref{27c}) yielding
$$
Z(k)={m_{\rm e}\over m_{\rm p}}
{b(|k|)\over a_{\rm p}(|R_{\rm p}k|^{-1})}[1-(kR_{\rm p})^{-2}]\ \times
$$
\begin{equation}
\sum_{\rm x=e,p} H[|k|-R_{\rm x}^{-1}]\int_{-1}^{{-{\rm sgn}(\Omega_{\rm x,0})}\over {R_{\rm x}k}}d\mu\ [F_{\rm x}(\mu ,t)-F_{\rm x}(\mu ,0)]  \label{34} 
\end{equation}
The system of the coupled equations (\ref{27a},\ref{27b},\ref{27c})
and (\ref{34}) describes the temporal
development of the particle distribution functions $F_{\rm p}$ und $F_{\rm e}$ under
the influence of self-excited Alfv\'en waves propagating either forward (+)
or backward (-). Here the boundary conditions are as follows: in the beginning
($t=0$) there is the mono-energetic beam distribution (\ref{1}), i.e.
in terms of the normalised distribution (\ref{21})
\be
F_{\rm p,e}(\mu ,t=0)=\delta (\mu +1) \ \label{35}\ee
The final state of the isotropisation phase is reached 
at time $T_i$ when both growth rate and temporal
derivative of the distribution disappear, i.e. when 
$\partial F_i/\partial \mu =0$. Consequently
\be
F_{\rm p,e}(\mu , t=T_i)={1\over 2} \label{36}\ee
At this time the Alfv\'en waves have completely isotropised the beam distrution.
In order to estimate this time scale $T_i$ or the associated isotropisation
length $\lambda $ we consider the extreme case that the wave spectrum is
constant and equals the wave intensity spectrum after the isotropisation.
By this method an approximation of the pitch angle Fokker-Planck 
coefficient $D_{\rm \mu\mu }$ 
is possible so that one gets a strict lower limit to the isotropisation
length $\lambda $. By demonstrating that this length is much smaller than the 
thickness $d$ of the blast wave region we will establish that the inflowing
proton-electron-beam is effectively iso\-tro\-pised in the blast wave plasma.

Inserting Eqs. (\ref{35}) and (\ref{36}) we find 
with
\be
\int_{-1}^{-1/R_{\rm p}k}d\mu [F_{\rm p}(\mu ,T_i)-F_{\rm p}(\mu ,t=0)]
=-{1\over 2}\bigl(1+{1\over R_{\rm p}k}\bigr)\ \ee
and
\be
\int_{-1}^{1/R_{\rm e}k}d\mu [F_{\rm e}(\mu ,T_i)-F_{\rm e}(\mu ,t=0)]
=-{1\over 2}\bigl(1-{1\over R_{\rm e}k}\bigr)\ \ee
that Eq. (\ref{34}) becomes
\begin{eqnarray}
Z(k)&=&-{m_{\rm e}\over m_{\rm p}}
{b(|k|)\over 2a_{\rm p}(|R_{\rm p}k|^{-1})}[1-(kR_{\rm p})^{-2}]\hfill\qquad\nonumber\\
 & &\times \sum_{\rm x=e,p}\,[1+(kR_{\rm x})^{-1}]H[|k|-R_{\rm x}^{-1}]
\hfill\qquad\nonumber \\
 &=&-{m_{\rm e}\over m_{\rm p}}{V_{\rm A}\over 2v}{\omega_{\rm p,e}^2\over c^2k^2}
B^2R_{\rm p}{n_{\rm i}\over n_{\rm b}\Gamma }\hfill\qquad\nonumber\\
 & &\times\sum_{\rm x=e,p}\,[1+(kR_{\rm x})^{-1}]H[|k|-R_{\rm x}^{-1}]
\hfill\qquad\nonumber\\
&\simeq& -Z_0k^{-2}
\label{38}\end{eqnarray}
The general solution of Eqs. (\ref{27a}) and (\ref{27b}) at time $t=T_i$ is
\begin{eqnarray}
I_+(T_i)=&\sqrt {Y+{1\over 4}(Z+I_+(0)-I_-(0))^2}&\hfill\qquad\nonumber\\
&+0.5\,\left(Z+I_+(0)-I_-(0)\right)&\hfill\qquad\label{39a}\end{eqnarray}
\begin{eqnarray}
I_-(T_i)=&\sqrt {Y+{1\over 4}(Z+I_+(0)-I_-(0))^2}&\hfill\qquad\nonumber\\
&-0.5\,\left(Z+I_+(0)-I_-(0)\right)&\hfill\qquad\label{39b}\end{eqnarray}
where
\be
Y\equiv I_+(0)I_-(0) \label{39c}\ee
If the initial turbulence is much weaker than the self-generated turbulence
$I(k,0)<<|Z(k)|$ and has a vanishing cross-he\-li\-ci\-ty $I_+(k,0)=I_-(k,0)=I(k,0)$
we obtain for Eqs. (\ref{39a},\ref{39b},\ref{39c}) approximately
\be
I_{\pm }(k,T_i)\simeq {1\over 2}[|Z|\pm Z]+\; {I^2(k,0)\over |Z|}\label{40}\ee
According to Eq. (\ref{38}) $Z(k)$ is negative so that Eq. (\ref{40}) reduces to
\be
I_+(k,T_i)\simeq {I^2(k,0)\over |Z(k)|} \label{41a}\ee
and
\be
I_-(k,T_i)\simeq |Z(k)| \label{41b}\ee
i.e. the
beam mainly generates backward moving Alfv\'en waves in the blast wave plasma.

\subsection{Energy budget}

The total enhancement in magnetic field fluctuation power due to 
proton and electron isotropisation is obtained by integrating Eq. (\ref{27b})
using Eq. (\ref{38})
\begin{eqnarray}
(\delta B_-)^2& =& \int_{-\infty}^\infty dk\bigl[
I_-(k,T_i)-I_-(k,0)]\nonumber\ \hfil\\
&=&-\int_{-\infty}^\infty dk\,Z(|k|)\nonumber\ \hfil\\
&=&{m_{\rm e}\over m_{\rm p}} {V_{\rm A}\omega_{\rm p,e}^2\over vc^2}
B^2R_{\rm p}^2{n_{\rm i}\over n_{\rm b}\Gamma }
[1+({m_{\rm e}\over m_{\rm p}})]\nonumber \quad\hfill\\
&\simeq& 4\pi V_{\rm A} n_{\rm i}m_{\rm p}v\Gamma [1+({m_{\rm e}\over m_{\rm p}})]\qquad\hfill\label{42}
\end{eqnarray}
so that the change in the magnetic field fluctuation energy density is
\be
\Delta U_{\delta B}={(\delta B_-)^2\over 8\pi }=
{1\over 2}V_{\rm A} n_{\rm i}m_{\rm p}v\Gamma [1+({m_{\rm e}\over m_{\rm p}})]\ \ee
Alfv\'en waves possess equipartition of wave energy density between magnetic and
plasma velocity fluctuations, so that the total change in 
fluctuation energy density due to pitch angle isotropisation
is
\be
\Delta U=\Delta U_{\delta v}+ \Delta U_{\delta B}=
2\Delta U_{\delta B}=V_{\rm A} n_{\rm i}m_{\rm p}v\Gamma [1+{m_{\rm e}\over m_{\rm p}}]\ \label{44}\ee
For consistency we show that this increase in the energy density of the
fluctuations is balanced by a corresponding decrease 
in the energy density of the beam protons and electrons during their
isotropization. We follow here the argument of
Bogdan et al. (\cite{bls91}), made originally for pick-up ions in
the solar wind, and generalise it to relativistic beam velocities.

As the beam particles scatter away from their initial pitch angle 
$\alpha =\pi $ and speed $V$, at each intermediate $\mu =p_{\parallel}/p$
they are confined to scatter approximately on a sphere
centered on the average wave speed of waves which resonate with that pitch
angle, given by (Skilling \cite{ski75}, Schlickeiser \cite{sch89})
\be
V_{\rm ph}=V_{\rm A}{I_+((R_i\mu )^{-1})-I_-((R_i\mu )^{-1})\over 
I_+((R_i\mu )^{-1})+I_-((R_i\mu )^{-1})}\simeq -V_{\rm A}\ \ee
because according to Eqs.~(\ref{41a},\ref{41b}) the intensity of backward
moving waves is much larger than the intensity of forward
moving waves. Thus the beam particles scatter in momentum space
on average onto a surface axisymmetric about $\vec{e}_z$ bounded by a
sphere centered at $-V_{\rm A}$, and passing through the ring of beam particle
injection. Once the beam particles are uniformly distributed over
the surface $V_{\rm f}(\mu )$, their energy density $w_T$ is
\be
w_{\rm T}=\sum _i\int d^3p\, f_i(p, \mu, t=t_i)\ \gamma_i m_ic^2\ \label{46}\ee
where the surface $V_f(\mu )$ is the relativistic sum of the initial velocity $V$ and 
\be
V_2=\int _{\cos (\alpha )}^{\mu }d\mu ^{´}V_{\rm ph}(\mu ^{´})
=-V_{\rm A}(\mu +1)\ \ee
We readily find
\be
V_{\rm f}(\mu )={V+V_2\over 1+{VV_2\over c^2}}\ \ee
or in units of the speed of light $c\ $ ($\beta =V/c$,
$\beta _{\rm f}(\mu )=V_{\rm f}/c$, $\beta_{\rm A}=V_{\rm A}/c$) we obtain
\be
\beta _{\rm f}(\mu )={\beta -\beta_{\rm A}(\mu +1)\over 1-\beta \beta_{\rm A}(\mu +1)}
\ \ee
The phase space distribution function of the
beam particles after isotropisation is
\be
f_i(p,\mu ,t=t_I)={n_{\rm i}\over 4\pi p_{\rm f}^2}\delta \bigl(p-p_{\rm f}(\mu )\bigr)
\label{49}\ee
with $p_{\rm f}=m_ic\beta_{\rm f}\gamma_{\rm f}$ and $\gamma_{\rm f}=(1-\beta_{\rm f}^2)^{-1/2}$.
Inserting Eq. (\ref{49}) in Eq. (\ref{46}) we obtain
\begin{eqnarray}
w_I&=&\sum _i{n_{\rm i}m_ic^2\over 2}\int_{-1}^1d\mu \gamma_{\rm f}(\mu )\qquad\hfill\nonumber \\
&=&\sum _i{n_{\rm i}m_ic^2\over 2}\int_{-1}^1d\mu \qquad\hfill\nonumber \\
 & &{1-\, \beta \beta_{\rm A}(\mu +1)\over 
\sqrt{[1-\beta \beta_{\rm A}(\mu +1)^2-[\beta -\beta (\mu +1)]^2}}
\end{eqnarray}
Substituting $u=\mu +1$ we derive
\begin{eqnarray}
w_{\rm I}&=&\sum _i{n_{\rm i}m_ic^2\over 2}\Gamma \int_0^2du {1-\beta \beta_{\rm A}\over 
\sqrt{1-\beta_{\rm A}^2u^2}}\qquad\hfill\nonumber \\
&=&\sum _i{n_{\rm i}m_ic^2\Gamma \over {2\beta_{\rm A}}}\Bigl[\arcsin (2\beta_{\rm A})
-\beta+\beta\sqrt{1-4\beta_{\rm A}^2}\Bigr]\label{51}
\end{eqnarray}
For our magnetic field strength $B\simeq $1 G and blast wave densities
$n_{\rm b}\simeq 10^6n^*_{\rm b,8}/\Gamma _{100}$ cm$^{-3}$, the Alfv\'en 
speed $\beta_{\rm A}\simeq 10^{-2}$
is much less than unity, so that we may expand Eq. (\ref{51}) for small values of $\beta_{\rm A}$ yielding
\begin{eqnarray}
w_{\rm I}&\simeq &\sum _in_{\rm i}m_ic^2\Gamma 
\Bigl[1+{2\over 3}\beta_{\rm A}^2+{6\over 5}\beta_{\rm A}^4-\, \beta \beta_{\rm A}(1-
\beta_{\rm A}^2)\Bigr]\qquad\hfill\nonumber \\
&\simeq &\sum _in_{\rm i}m_ic^2\Gamma [1-\beta_{\rm A}\beta ]\label{52}\end{eqnarray}
The initial energy density of the beam particles $w_0$ can be calculated
using the initial beam distribution function (\ref{1}), i.e. 
\be
f_i(p,\mu ,t=0)={n_{\rm i}\over 2\pi p^2}\delta (\mu +1)
\delta \bigl(p-P\bigr)
\ee
yielding
\be
w_0=
\sum _i\int d^3p\, f_i(p, \mu, t=0) \gamma_i m_ic^2
=\sum _in_{\rm i}m_ic^2\Gamma 
\label{54}\ee
Obviously the change in the energy density of the beam particles then is
\begin{eqnarray}
\Delta w&=&w_{\rm I}-w_0=-\sum _in_{\rm i}m_ic^2\Gamma \beta_{\rm A}\beta\qquad\hfill\nonumber \\
&=&-n_{\rm i}m_{\rm p}vV_{\rm A}\Gamma [1+{m_{\rm e}\over m_{\rm p}}]\label{55}\end{eqnarray}
which is exactly $-\Delta U$ from Eq. (\ref{44}). 
The plasma turbulence is generated at the expense of the beam particles
which relax to a state of lower energy density; or with other words,
the excess energy density in the non-isotropic beam distribution is
transferred to magnetohydrodynamic plasma waves that scatter the beam particles
to an isotropic distribution in the blast wave plasma. If this isotropisation
is quick enough -- which we will calculate next -- the beam particles
attain an almost perfect isotropic distribution function (\ref{49})
in the blast wave plasma, because to lowest order
in $\beta_{\rm A}$ the final particle momentum $p_{\rm f}\simeq \beta \Gamma m_ic$
is independent of $\mu $.

\subsection{Isotropisation length}
Using the fully-developed turbulence spectra (\ref{41a},\ref{41b}) in
Eq. (\ref{Dmumu}) we obtain for
the pitch angle Fokker-Planck coefficient of the beam particle $i$
\be
D_{\rm \mu\mu }={\pi \Omega _i^2\over 2}{1-\mu ^2\over B^2} \; d_{\rm \mu \mu }
\label{56}\ee
where 
\begin{eqnarray}
d_{\rm \mu \mu }&=&\sum_{\pm }[1\mp {\mu V_{\rm A}\over v}]^2
{I_{\pm }\bigl({\Omega _i\over v\mu \mp V_{\rm A}}\Bigr)\over |\mu v\mp V_{\rm A}|} \qquad\hfill\nonumber \\
&\simeq& [
{I_+\bigl({\Omega _i\over v\mu - V_{\rm A}}\Bigr)\over |\mu v- V_{\rm A}|}+\;
{I_-\bigl({\Omega _i\over v\mu + V_{\rm A}}\Bigr)\over |\mu v+ V_{\rm A}|}]
\qquad\hfill\nonumber \\
&=&
{|Z_0||v\mu +V_{\rm A}|\over \Omega _i^2}+\; 
{I^2(\Omega _i/(v\mu -V_{\rm A}),0)\Omega _i^2\over 
|Z_0||v\mu -V_{\rm A}|^3}\label{57}\end{eqnarray}
For ease of exposition we assume that the initial turbulence spectrum
has the form $I(k,0)=I_0k^{-2}$, so that Eq. (\ref{57}) becomes
\be
d_{\rm \mu \mu }={|Z_0|\over \Omega _i^2}\Bigl[|v\mu +V_{\rm A}|+\Delta 
|v\mu -V_{\rm A}|\Bigr]\label{58}\ee
where $\Delta \equiv (I_0/|Z_0|)^2$.

According to quasilinear theory (e.g. Earl \cite{earl73}, Schlickeiser 
\cite{sch89}) as a consequence of pitch-angle scattering 
the beam particles adjusts to the isotropic distribution (\ref{36})
on a length scale given by the scattering length 
\be
\lambda ={3v\over 8}\int_{-1}^1d\mu {(1-\mu ^2)^2\over D_{\rm \mu\mu }(\mu )}
={3\over 4\pi }{B^2\over |Z_0|}J(\epsilon , \Delta )\label{59}\ee
with $\epsilon =V_{\rm A}/v$ and
\be
J(\epsilon ,\Delta )\equiv \int _{-1}^1d\mu (1-\mu ^2)
\bigl[|\mu +\epsilon |+ \Delta |\mu -\epsilon |\bigr]^{-1}\label{60}\ee
To obtain Eq. (\ref{59}) we have inserted Eqs. (\ref{56}) and (\ref{58}).
After staightforward
but tedious integration the value of the integral (\ref{60}) to lowest order
in the small parameters $\epsilon <<1$ and $\Delta <<1$ is
\be
J(\epsilon ,\Delta )\simeq -2\ln (\sqrt{2}\epsilon \Delta )\label{61}\ee
so that the scattering length (59) becomes with Eq. (\ref{38})
\begin{eqnarray}
\lambda &\simeq &{3\over 2\pi }{B^2\over |Z_0|}[-\ln (\sqrt{2}\epsilon \Delta )]
\qquad\hfill\nonumber \\
&=&{3\over \pi }{m_{\rm p}\over m_{\rm e}}{c^2\over \omega_{\rm p,e}^2R_{\rm p}}{v\over V_{\rm A}}
{n_{\rm b}\Gamma\over n_{\rm i}}[-\ln (\sqrt{2}\epsilon \Delta )]\qquad\hfill\nonumber \\
&=&{3\over \pi }
{c\over \omega _{\mathrm p,i}}{n_{\rm b}\over n_{\rm i}}\ln (\sqrt{2}\epsilon \Delta )^{-1}
\label{62}\end{eqnarray}
Inserting our typical parameter values we obtain 
\be
\lambda =2.2\cdot 10^{11} {n_{\rm b,8}^{1/2}\over n_{\rm i}}
\ln (\sqrt{2}\beta_{\rm A}\Delta )^{-1}\;\; \, \rm{cm}\label{63}\ee
Note that the ratio of initial to fully developed turbulence intensities
enters only weakly via the logarithm. For turbulence intensity ratios
from $10^{-1}$ to
$10^{-5}$ implying values of $\Delta $ from
$10^{-2}$ to $10^{-10}$ we find with $\beta_{\rm A}\simeq 
10^{-2}$ that
$\ln [\sqrt {2}\beta_{\rm A}\Delta ]^{-1}$ varies between
$8.9$ and $27.3$. Taking the larger value in Eq. (\ref{63}) 
yields for the scattering length in the blast wave plasma
\be
\lambda \simeq 6\cdot 10^{10}\; 
{n_{\rm b,8}^{1/2}\over \Gamma _{100}\,n^*_{\rm i}}\;\;
\; \rm{cm}\label{65}\ee
which corresponds to an isotropisation time scale of
\be
t_{\rm R}=\lambda /c=2 \; 
{n_{\rm b,8}^{1/2}\over \Gamma _{100}\,n^*_{\rm i}}\;\; 
\; \rm{s}\label{69}\ee
If the thickness $d$ of the blast wave region is larger than 
the scattering length (75), indeed an isotropic distribution
of the inflowing interstellar protons and electrons with Lorentz factor
(see Eq. (\ref{52})) $<\Gamma >=\Gamma (1-\beta_{\rm A}\beta )\simeq \Gamma $
in the blast wave frame 
is effectively generated. In the following
sections we investigate the radiation products 
resulting from the inelastic interactions of these primary particles
with the cold blast wave plasma.

This discussion of particle isotropization on self-excited turbulence
applies to both AGN and GRBs. The radiation modelling of GRBs often requires
energy equipartition between electrons and protons (Katz \cite{ka94}), 
based on detailed plasma physics considerations 
(e.g. Beloborodov \& Demia\'nski \cite{be95}; Smolsky \& Usov \cite{su96}; 
Smolsky \& Usov
\cite{su99}). However, the isotropization itself provides electrons with an energy roughly 2000 times smaller than that of the protons. Therefore a 
reacceleration of electrons would be required to reach equipartition.
Since the energy density of the turbulence is only a fraction 
($\propto \beta_{\rm A}$) of the energy density of the incoming beam, the 
turbulence is not energetic enough to reaccelerate electrons for
$\beta_{\rm A} \ll 1$, and further studies may be required to understand 
the production of radiation in GRBs. The situation is different with AGN,
for which equipartition is not required and for which the variability 
timescales are substantially longer. Here we will
discuss the radiation modelling for AGN on the basis of the particle
distributions resulting from the isotropization process alone.

\section{Radiation modelling of the blast wave}

Now we deal with isotropic particle distribution functions in the
blast wave frame, which itself is not stationary because the blast wave sweeps
up matter and thus momentum. Momentum conservation then requires a deceleration
of the blast wave depending on whether or not the swept-up
particles maintain their initial kinetic energy, and depending on possible
momentum loss from anisotropic emission of electromagnetic radiation.

As we have sketched in Fig.\ref{sketch}, 
the blast wave is assumed to have a disk-like geometry with constant radius
$R$  and thickness $d$ that moves with bulk Lorentz factor $\Gamma$.
The matter density in that disk $n_{\rm b}$ is supposed to be orders of magnitude
higher than that of the ambient medium $n_{i}^\ast$. 

In the blast wave frame the external density $n_{i} = \Gamma
\,n_{\rm i}^\ast$ and sweep-up occurs at a rate
\be
\dot N(\gamma) = \pi\,R^2\,c\,n_{\rm i}^\ast\,\sqrt{\Gamma^2 -1}\
\delta (\gamma-\Gamma)\ . \label{injec}
\ee
As we have seen in the preceding section, the relativistic electrons
and protons get very quickly isotropised and lose little energy in the 
process. This has two consequences: the sweep-up is a source of isotropic,
quasi-monoenergetic protons and electrons with Lorentz factor $\Gamma$
in the blast wave frame. The isotropisation also provides a momentum transfer
from the ambient medium to the blast wave. 

\subsection{The equation of motion of the blast wave}

In a time interval $\delta t$
the blast wave sweeps up a momentum of
\be
\delta \Pi = \pi\,R^2\,m_{\rm p}\,c^2\,n_{\rm i}^\ast\,(\Gamma^2 -1)\, \delta t
\ee
which is transferred from the swept-up particles to the whole system
with mass
\be
M_{\rm BW}= \pi\,R^2\,d\,\left(n_{\rm b}\,m_{\rm p} + n_{\rm nth}\, m_{\rm nth}\right)
\ee
where the density and relativistic mass of the energetic particles
\be
n_{\rm nth}\, m_{\rm nth} = m_{\rm p}\, \int_1^\infty \rm{d}\gamma\ \gamma\, n(\gamma)
\label{massload}\ee
have to be added to that of the thermal plasma. Therefore, the blast wave
will tend to move backwards and its Lorentz factor relative to the ambient
medium is reduced to
\be
\Gamma^\prime = \Gamma\,\sqrt{1+\left({{\delta \Pi}\over {M_{\rm BW}\,c}}\right)^2}
\,-\, \sqrt{\Gamma^2 -1}\,{{\delta \Pi}\over {M_{\rm BW}\,c}}
\label{decel}\ee
which can easily be integrated numerically.
For a highly relativistic blast wave we can expand the blast wave equation
of motion and derive the timescale for slowing down 
\be
\tau_{\rm BW} =-\,{{\Gamma}\over {{{\delta \Gamma}\over {\delta t}}}} 
\simeq
{{\Gamma\,d\,\left(n_{\rm b}\,m_{\rm p} + n_{\rm nth}\, m_{\rm nth}\right)}\over 
{n_{\rm i}^\ast\,m_{\rm p}\,c\,(\Gamma^2-1)^{3/2}}}\ .
\label{82}\ee
In the laboratory frame this timescale is longer by a factor $\Gamma$.

\begin{figure}
\resizebox{\hsize}{!}{\includegraphics{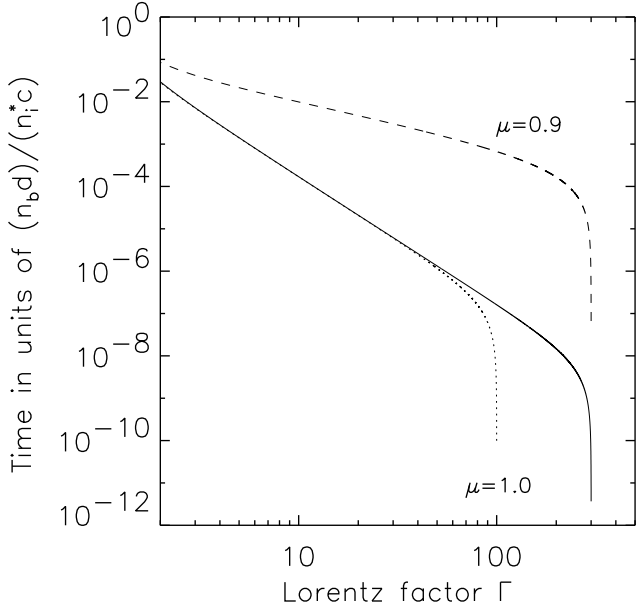}}
\caption{The observed time plotted versus the Lorentz factor of
the blast wave. The solid line is calculated for an observer located in
the direction of the jet, i.e. $\mu$=1, and initial Lorentz factor 300. 
The dotted line, derived for an initial Lorentz factor of 100, and the dashed line, calculated for an aspect cosine $\mu$=0.9, show that
the observed time needed for the deceleration to semirelativistic
velocities is independent of the initial
Lorentz factor and varies little with the aspect angle.}
\label{timeplot}
\end{figure}

We may also integrate the inverse of the slow-down rate to derive the
observed time frame of the deceleration, that is the time at which
the blast wave would be observed with a particular Lorentz factor.
This may be interesting to compare with the results of VLBI observations
of blazars. In the radiative regime, i.e. when the internal energy is
radiated away quickly, we may neglect the mass loading for $\Gamma\gg 1$ and 
obtain
\begin{eqnarray}
\tau^\ast &=& \int_{\Gamma_0}^{\Gamma} d\Gamma^\prime\ 
{{\Gamma^\prime\,(1-\beta^\prime\mu)}\over {{{\delta \Gamma}
\over {\delta t}}}}\nonumber\\
&=& {{n_{\rm b}\,d}\over {n_{\rm i}^\ast\,c}}\ 
\int_{\Gamma}^{\Gamma_0} d\Gamma^\prime\ 
{{\Gamma^\prime\,(1-\beta^\prime\mu^\ast)}\over {
(\Gamma^2-1)^{3/2}}}
\label{slow}
\end{eqnarray}
where we have assumed an interstellar medium with a constant density 
$n_{\rm i}^\ast$. In Fig.~\ref{timeplot} we show typical solutions of the
integral Eq.(\ref{slow}). The observed time needed for a deceleration
to Lorentz factors $\Gamma$ of a few is independent of the initial
Lorentz factor $\Gamma_0 \gg \Gamma$, and it varies much less strongly
with the aspect angle than did the initial Doppler factor.

We can also calculate the distance traveled by the blast wave. For this
we replace the retardation factor in Eq.(\ref{slow}) by $\beta c$
and we obtain for $\Gamma_0 \gg \Gamma$
\be
L={{n_{\rm b}\,d}\over {2\,n_{\rm i}^\ast}}\ \ln{{\Gamma +1}\over {\Gamma -1}}
\label{dista}
\ee
Comparing the swept-up mass with the initial mass of the blast wave we
see that mass loading in the radiative regime is important only for
$\Gamma\la 2$, and thus its neglect in the derivation of Eqs.~\ref{slow}
and \ref{dista} is justified.
If the blast wave is strong enough, $L$ will be in the range of a few
kpc
and the blast wave may leave the host galaxy at relativistic speed.
It may travel through the dilute intergalactic medium for
several Mpc, before it eventually runs into a denser gas structure and 
converts its kinetic energy into radiation. Thus the power of the
blast wave, i.e.
the luminosity, is linked to the probability of escape from the host
galaxy. When the systems are viewed from the side, they would fit
into the morphological classification of Fanaroff-Riley I and
Fanaroff-Riley II galaxies: while the former are less luminous, they are 
also core dominated, whereas the latter exhibit very luminous radio lobes.

\subsection{The evolution of the particle spectra}

Eq. (\ref{injec}) states the differential injection of relativistic
particles in the blast wave. Here we concentrate on the protons because
they receive a factor of $m_{\rm p} /m_{\rm e}$ more power than electrons, which also
have a low radiation efficiency for $\gamma \ll 1000$. Electrons
(and positrons) are supplied much more efficiently as secondary particles
following inelastic collisions of the relativistic protons. Since
no reacceleration is assumed, the continuity equations for protons
and secondary electrons read
\be
{{\partial N_{\rm p}(\gamma)}\over  {\partial t}} + {{\partial }\over  {\partial 
\gamma}} \Big(\dot \gamma\, N_{\rm p}(\gamma)\Big) +
{{N_{\rm p} (\gamma) }\over {T_{\rm E}}}+
{{N_{\rm p} (\gamma) }\over {T_{\rm N}}}=\dot N_{\rm p}(\gamma) 
\label{contp}
\ee
\be
{{\partial N_{\rm e}(\gamma)}\over  {\partial t}} + {{\partial }\over  {\partial 
\gamma}} \Big(\dot \gamma\, N_{\rm e}(\gamma)\Big) +
{{N_{\rm e} (\gamma) }\over {T_{\rm E}}}+
{{N_{\rm e} (\gamma) }\over {T_{\rm ann}}}=\dot N_{\rm e}(\gamma)
\label{conti}
\ee
where $T_{\rm E}$ is the timescale for diffusive escape, and 
$T_{\rm N}$ is the 
loss timescale associated with $(p\rightarrow n)$ reactions with neutron decay occurring outside the blast wave.
For positrons catastrophic annihilation losses on a timescale
$T_{\rm ann}$ are taken into account. 
The injection rate of secondary
electron is related to the rate of inelastic collisions.

Energy losses arise from elastic scattering at a level of 
\begin{equation}
- \dot \gamma_{\rm el,p} =
3\cdot 10^{-16}\,n_{\rm b}\, {{\gamma}\over {\sqrt{\gamma^2 -1}}}
\end{equation}
where we have assumed the ambient medium as a mixture of 90 \% hydrogen
and 10 \% helium.
The energy losses caused by inelastic collisions can be determined
by integration over the
secondary particle yield. Here we use the Monte Carlo model DTUNUC (V2.2)
(M\"ohring \& Ranft \cite{MR91}, Ranft et al. 
\cite{RCT94}, Ferrari et al. \cite{Fer96a}, Engel et al. 
\cite{ERR97}),
which is based on a dual parton model (Capella et al. \cite{Cap94}).
This MC model for hadron-nucleus and nucleus-nucleus interactions
includes various modern aspects of high-energy physics and has been
successfully applied to the description of hadron production in
high-energy collisions (Ferrari et al. \cite{Fer96b}, Ranft \& Roesler
\cite{RR94}, M\"ohring et al. \cite{Moe93}, Roesler et al.
\cite{RER98}). The total energy
losses from inelastic collisions are well approximated by
\begin{equation}
- \dot \gamma_{\rm inel} =
7\cdot 10^{-16}\,n_{\rm b}\, {{(\gamma-1)^2}\over {\gamma+1}}
\end{equation}
The timescale for neutron escape after $(p\rightarrow n)$ reactions
at relativistic energies is approximately
\be
T_{\rm N} \simeq 3\cdot 10^{15}\ n_{\rm b}^{-1}\ \left\langle\exp\left(-{L\over {
\gamma\,2.7\cdot 10^{13}\ {\rm cm}}}\right)\right\rangle\quad {\rm sec}
\label{T_N}
\ee
where $L$ is the neutron path length through the blast wave and the brackets
denote an average over the neutron emission angle. In all the examples 
discussed in this paper the exponential
in Eq. (\ref{T_N}) will be close to unity.

The Monte-Carlo code also provides the differential cross sections
for the pion production, on which we base our calculation of the pion
source functions. Neutral pions decay immediately into two \grs and the
\gr source function
\be
q_\gamma (\epsilon) = 2\int_{\epsilon + {{m\pi^2}\over {4\epsilon}}}^\infty 
dE_\pi\ {{q_{\pi^0}}\over \sqrt{E_\pi^2 - m_{\pi^0}^2}}
\ee
Charged pions decay into muons and finally into electrons or positrons. The
secondary electron source functions are essentially determined by the kinematic
of the two decay processes (Jones \cite{jon63}, Pohl \cite{poh94}).
The secondary electrons loose energy mainly by inverse Compton 
scattering, synchrotron emission, brems\-strahlung, and elastic scattering.
\begin{eqnarray}
-\dot \gamma_{\rm IC}&=&2.7\cdot 10^{-14}\,{{U_{\rm ph}}\over {m_{\rm e} c^2}}\ 
\left(\gamma^2 -1\right) \\
-\dot \gamma_{\rm Sy}&=&2.7\cdot 10^{-14}\,{{U_{\rm B}}\over {m_{\rm e} c^2}}\ 
\left(\gamma^2 -1\right) \\
-\dot \gamma_{\rm Br}&=& 8\cdot 10^{-16}\ n_{\rm b}\, (\gamma-\gamma^{-1}) \\
-\dot \gamma_{\rm el,e}&=&
6\cdot 10^{-13}\,n_{\rm b}\, {{\gamma}\over {\sqrt{\gamma^2 -1}}}
\end{eqnarray}
where the energy densities $U_{\rm B}$ and $U_{\rm ph}$ are in units of
${\rm eV/cm^3}$.
The probability of annihilation per time interval is (Jauch \& Rohrlich
\cite{jr76})
\begin{eqnarray}
T_{\rm ann}^{-1}&= & {{3\,c\,\sigma_{\rm T} \,n_{\rm b}}\over
{8\,(\gamma +1)\sqrt{\gamma^2 -1}}}\ \times  \\
 & &\left[ \left(\gamma +4 +\gamma^{-1}\right)
\ln \left(\gamma + \sqrt{\gamma^2 -1} \right) - \beta \,(\gamma +3)\right]
\nonumber
\end{eqnarray}
where we have assumed that the temperature of the background plasmas is higher
than about 100 eV, so that a Coulomb correction and contributions from
radiative recombination can be neglected.

The diffusive escape of particles from the blast wave scales
with the scattering length
$\lambda$ as derived in Eq.(\ref{65}). For the escape timescale we can write
\be
T_{\rm E} = {{d^2}\over D} = {{3\,d^2}\over {\lambda\,\beta\,c}} \simeq
1.67\cdot 10^{-19}\ {{d^2\, \Gamma\,n_{\rm i}^*}\over {\beta\,\sqrt{n_{\rm b}}}}
\quad{\rm sec}
\ee
If escape is more efficient than the energy losses via
pion production, then the bolometric luminosity of the blast wave will
be reduced by a factor $T_{\rm E}/\tau_\pi \simeq T_{\rm E}\,\vert
\dot \gamma_{\rm inel}\vert/\gamma$.

The proton spectra evolve differently depending on the relative effect
of the losses and the blast wave slow down. If there were no
losses, the relativistic mass loading Eq.~(\ref{massload}) would be trivial to calculate. Combining the mass loading equation with the deceleration equation Eq.~(\ref{decel}) we obtain after some calculus in the limit
$\Gamma \gg 1$
\be
\dot \Gamma = {{n_{\rm i}^\ast\,c}\over {n_{\rm b} \,d}} {{\Gamma^4}\over {\Gamma_0}}
\label{galoss}
\ee
where $\Gamma_0$ is the initial Lorentz factor of the blast wave.
By comparison with the simple timescale Eq.~(\ref{82}) we see that the mass
loading reduces the deceleration by a factor $\Gamma/\Gamma_0$. The proton
continuity equation (\ref{contp}) can be trivially integrated in the absence
of losses to
\begin{displaymath}
N(\gamma) =\int_0^t d\tau\ \dot N_{\rm p} (\gamma)
=C\, \int_\Gamma^{\Gamma_0} d\Gamma^\prime\ {{\dot N_{\rm p} (\gamma)}\over {{\Gamma^\prime}^{-4}}}
\end{displaymath}
\be
\hphantom{N(\gamma)}\simeq C^\prime\, \gamma^{-3}\qquad {\rm for}\ 
\Gamma \le \gamma\le\Gamma_0 
\ee
So if the energy losses can be neglected,
the blast wave deceleration alone would cause approximately a
$\gamma^{-3}$ spectrum in the relativistic range.

\begin{figure*}
\resizebox{11cm}{!}{\includegraphics{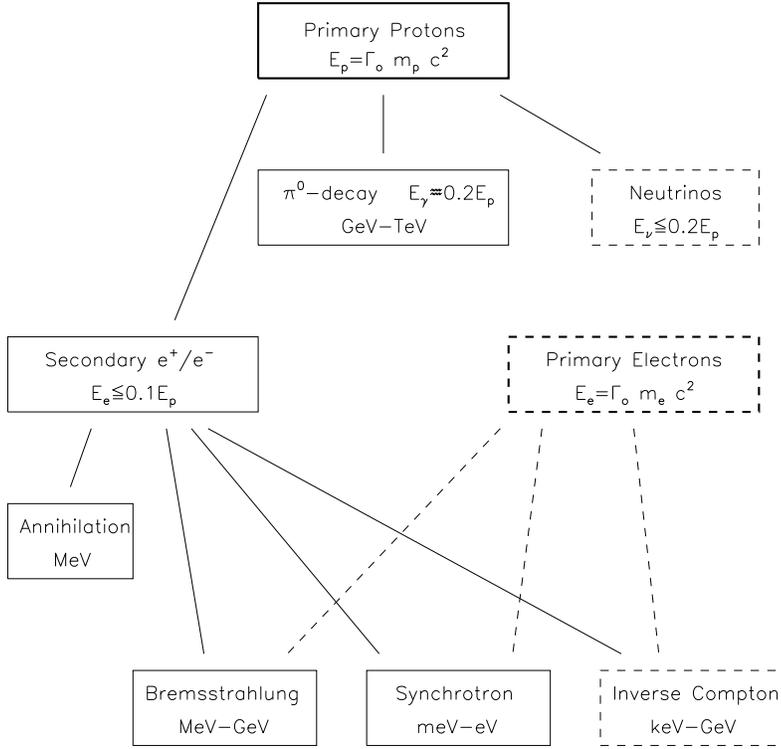}}
\hfill
\parbox[b]{65mm}{
\caption{Scheme of the reaction and radiation channels of relativistic
blast waves. Here the logical flow is from top to bottom.
In each box we also note the approximate energy of the particles
and the photons in the blast wave frame. Depending on the Doppler factor, 
i.e. on the aspect angle, the observed energy may be substantially higher.
The channels displayed in solid line boxes will be discussed
in detail in this paper, whereas those printed in dashed line boxes will
be the subject of forthcoming publications.}
\label{schema}}
\end{figure*}

\subsection{Optical depth at \gr energies}

Here we discuss a system of rather peculiar geometry, a very thin disk,
and a strong angular variation of optical depth effects can be expected.
While for photons traveling exactly in the blast wave plane the
line-of-sight through the system has an average length $R$, and the system
may appear optically thick, the angle-averaged line-of sight has a length 
\be
{\bar L} = {{d}\over 2}\ \ln\left(1+{{2\,R}\over d}\right)
\ee
which is approximately equal to $d$. Therefore the total optical depth
may be small because the photons would interact
only in a very small solid angle element. In all the cases presented here
this is true for the Thompson optical depth $\tau =\sigma_{\rm T}\,
n_{\rm e}\,
{\bar L}$, so that Compton scattering will not effectively influence
the emission, though $\sigma_{\rm T}\,n_{\rm e}\,R$ may be of the order of 1.

At photon energies above 511 keV in the blast wave frame photon-photon
pair production may occur and inhibit the escape of \grs from the system.
This process is not important in our model, because very high apparent
luminosities can be produced with moderate photon densities in the source
as a result of the highly relativistic motion of the blast wave and the 
associated high Doppler factors. For example, given a system radius
$R=10^{15}$ cm and a Doppler factor $D=100$, an apparent luminosity
$L=10^{48}\ {\rm erg\, sec^{-1}}$ at MeV energies in the blast wave rest 
frame corresponds to an optical depth of
$\tau_{\gamma\gamma}\simeq 10^{-4}$.

\section{High energy emission from the blast wave}

In Fig.~\ref{schema} we show the reaction and radiation channels 
of the swept-up particles together with the approximate photon
energies for the emission processes. Since the source power of
secondary electrons and the power emitted in the form of
$\pi^0$-decay \grs is the same, the bulk of the bolometric
luminosity will be emitted in the GeV to TeV energy range, independent
of the choice of parameters. Thus the unexpected finding, that 
\gr emission of blazars generally provides a major part of the 
bolometric luminosity, would find a natural explanation if
as in this model pion production in inelastic collisions were 
the main source of energetic electrons.

In this paper
we concentrate on the high-energy $\gamma$-ray emission, and especially
on the processes of $\pi^0$ decay, brems\-strah\-lung, and pair annihilation,
for they scale only with internal parameters like the gas density $n_{\rm b}$.
A discussion
of inverse Compton scattering and neutrino emission
will be presented elsewhere. The inverse Compton process
must inevitably occur, but its rate, the spectral and angular 
distribution depend on the actual choice of target photon field. If
this scattering provides significant energy losses for
electrons and positrons, the solution of Eq.~(\ref{conti}) will be 
particularly arduous.

The $\pi^0$ production spectrum
is calculated using the Monte-Carlo code DTUNUC
described earlier in this paper. 

The bremsstrahlung spectrum is determined using the relativistic limit
of the differential cross section (Blumenthal \& Gould \cite{bg70})
\be
{{d\sigma}\over {d\epsilon}}\Big\arrowvert_{BS}
 = {{\alpha\,r_{\rm e}^2}\over {\epsilon}}\,\phi_{\rm u}
\left[{4\over 3}-{4\over 3}{\epsilon\over \gamma}+
{{\epsilon^2}\over {\gamma^2}}\right]
\ee
where
$$\phi_{\rm u} = 4\,\ln\left[2\gamma\left({{\gamma}\over \epsilon}-1\right)
\right] \,-\,2 $$
and the photon energy $\epsilon$ is in units of $m_{\rm e}\,c^2$.
The annihilation spectrum is obtained from the differential cross section
(Jauch \& Rohrlich \cite{jr76})
\be
{{d\sigma}\over {d\epsilon}}\Big\arrowvert_{\rm AN} = 
{{\pi\,r_{\rm e}^2}\over {\beta^2\gamma^2}}
\left[{\epsilon\over {\epsilon_1}}+{{\epsilon_1}\over {\epsilon}}
+{{2(1+\gamma)}\over {\epsilon\epsilon_1}}-
\left({{1+\gamma}\over {\epsilon\epsilon_1}}\right)^2\right]
\ee
with 
$$\epsilon+\epsilon_1=1+\gamma\ \ {\rm and}\ \ {1\over {\gamma
(1+\beta)}} \le {\epsilon\over {\epsilon_1}}\le
\gamma (1+\beta)$$
We follow the evolution of the electron and positron spectra to Lorentz factors
of 3, below which the bremsstrahlung emissivity
is not calculated and the annihilation
spectrum is derived in the non-relativistic limit.

Synchrotron emission can be expected in the optical 
to X-ray frequency range. For head-on jets synchotron emission may be
observable at X-ray energies. The peak of the observed synchrotron spectra in
$\nu F_\nu$ representation scales roughly as
\begin{equation}
E_{\rm Sy,max} \simeq \left({{B}\over {\rm Gauss}}\right)\left({{\Gamma}
\over {200}}\right)^2\left({{D}
\over {200}}\right)\quad {\rm keV}
\label{symax}
\end{equation}
with $D$ denoting the Doppler factor,
but free-free absorption and 
the Razin-Tsytovich effect, in addition to synchrotron self-ab\-sorp\-tion,
would inhibit strong synchrotron emission in
the near infrared and at lower frequencies.
Radio emission may become observable at later phases, when the blast wave
has decelerated and the Doppler factor is reduced, or when the
blast wave medium is diluted, e.g. as a result of imperfect collimation.

The peak of the observed $\pi^0$-decay spectrum in $\nu F_\nu$
representation scales as
\begin{equation}
E_{\pi^0,max} \simeq \left({{\Gamma}
\over {100}}\right)\left({{D}
\over {100}}\right)\quad {\rm TeV}
\label{pimax}
\end{equation}
Comparing Eqs.(\ref{symax}9 and (\ref{pimax}) we see that hard X-ray synchrotron
emission as apparently observed from the BL Lacertae object Mkn 501 
(Pian et al. \cite{pia98}) implies that the $\nu F_\nu$ peak of the high energy
emission is located far in the TeV range of the spectrum. 
At this point we like to issue a warning to the reader, that one has
to be careful
when comparing the model spectra to actual data.
The TeV \gr spectra of real sources,
even the closeby ones, 
are modulated during the passage through the intergalactic medium, for 
the \gr photons undergo pair production in collision with infrared
background photons. Unfortunately the intergalactic infrared
background spectra, and thus the magnitude and the energy dependence 
of the optical depth, are not well known. 
The current limits are compatible with severe absorption
at all energies above 1 TeV even for Mkn 421 and Mkn 501. Therefore,
model spectra which display a $\nu F_\nu$ peak at 10 TeV are not incompatible
with the observed peak energies in the range of $\sim$ 0.5 TeV for
Mkn 421 and Mkn 501.

\begin{figure*}
\resizebox{\hsize}{!}{\includegraphics{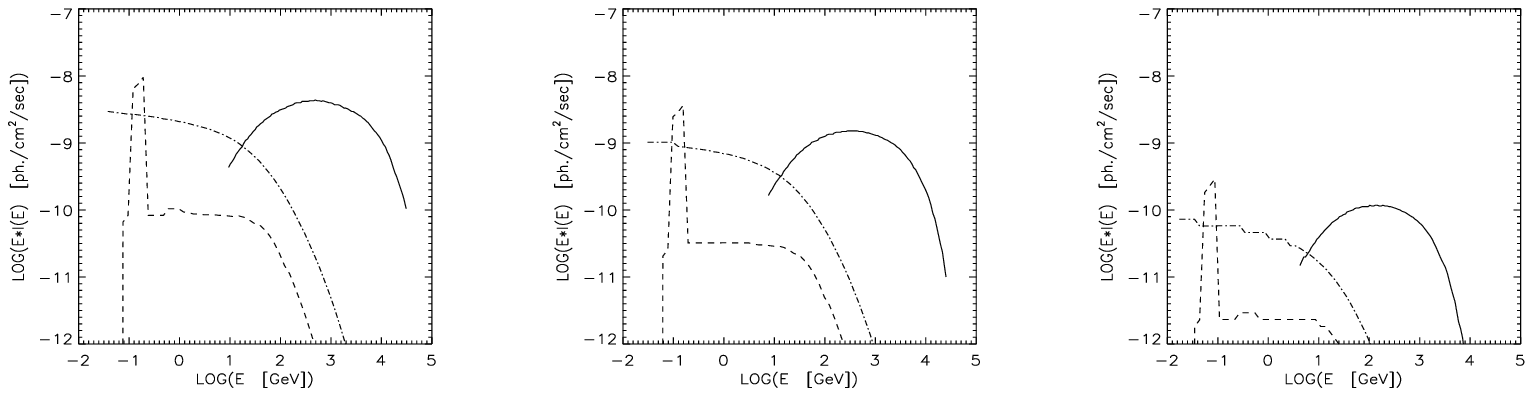}}
\resizebox{\hsize}{!}{\includegraphics{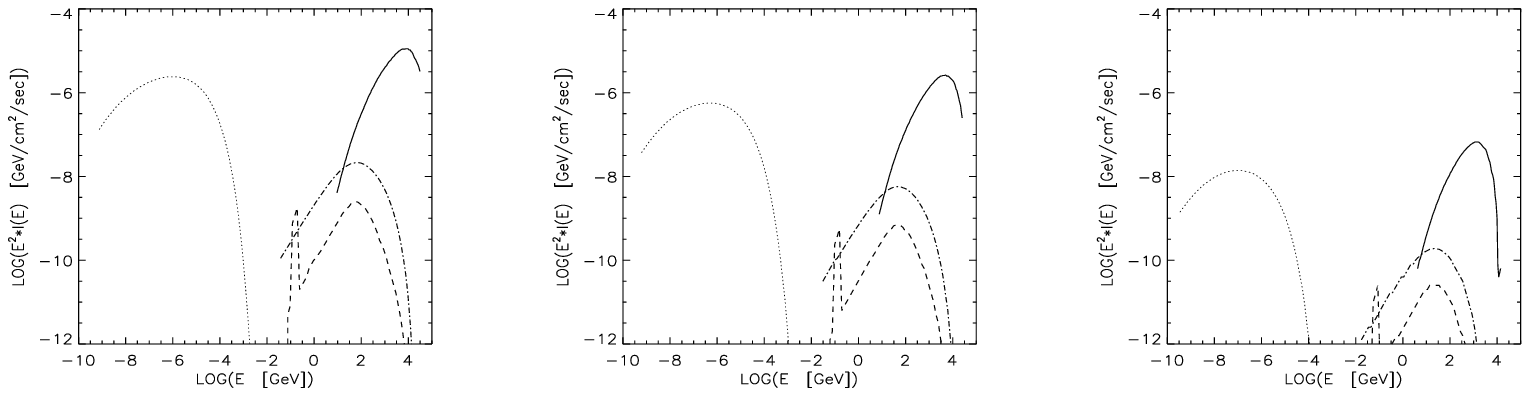}}
\caption{Spectral evolution of a relativistic blast wave in an
environment of constant density. The top row displays $F_\nu$ spectra of
the \gr emission, whereas the bottom row shows $\nu F_\nu$ spectra
from the optical to high energy \grs including also the synchrotron emission.
The solid lines display
$\pi^0$-decay $\gamma$-rays, the dot-dashed lines 
bremsstrahlung, the dashed lines annihilation emission, and the dotted
lines represent the synchrotron emission. From left to
right the panels show the spectra after one hour, 10 hours,
and 100 hours
observed time. The parameters are: $\Gamma_0$=300, d=$3\cdot 10^{13}$cm, 
R=$10^{14}$cm, B= 2.0 G, $n_{\rm i}^\ast$=0.2 cm$^{-3}$, $n_{\rm b}$=$5\cdot
10^8$  cm$^{-3}$, for an AGN at z=0.5 viewed at an angle $\theta_{\rm obs}=
0.1^\circ$. After 100 hours the blast wave has decelerated to $\Gamma\simeq 
106$.}
\label{speca}
\end{figure*}
\begin{figure*}
\resizebox{\hsize}{!}{\includegraphics{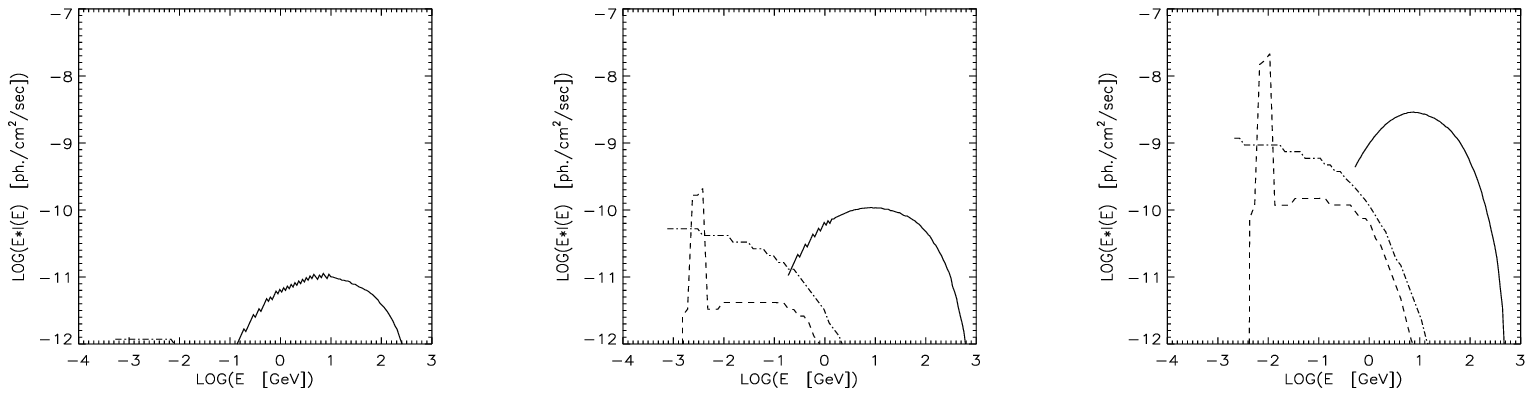}}
\resizebox{\hsize}{!}{\includegraphics{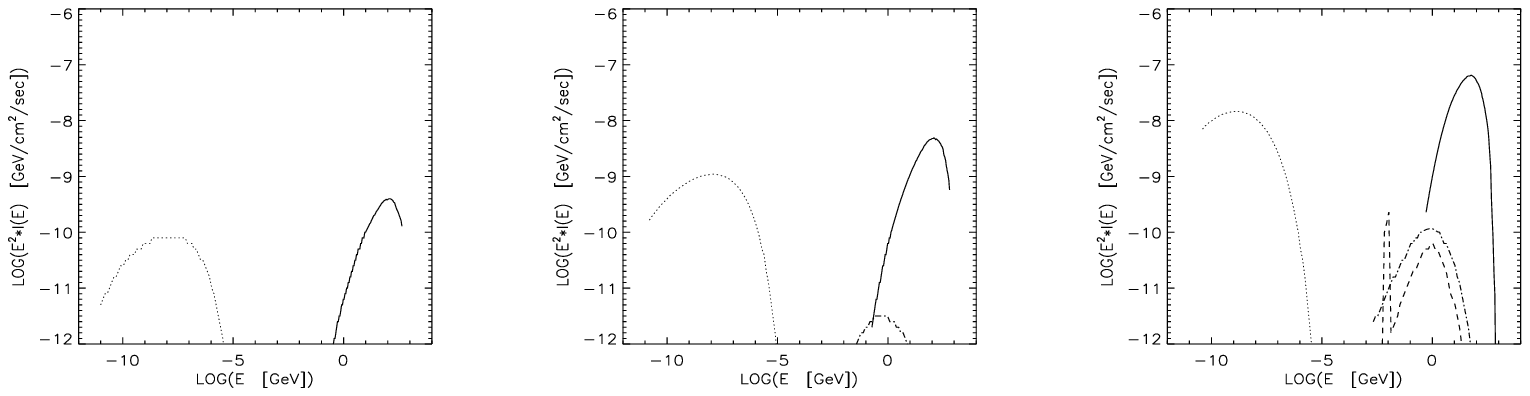}}
\caption{The same system as in Fig.~\ref{speca} viewed by an
observer located $\theta=2\degr$ off the blast wave direction of motion
($\mu$=0.99939).
Note that now the panels show the spectra after 10 hours, 100 
hours, and 1000 hours of observed time.
Here also d=$10^{14}$ cm, R=$2\cdot 10^{15}$ cm, 
$n_{\rm i}^\ast$=1.5 cm$^{-3}$, and $n_{\rm b}$=$10^8$ cm$^{-3}$
to make the source appear as bright as before. The system appears to
evolve much slower than for $\theta=0.1\degr$. Note that the left column here 
represents the same observed time as the middle column in Fig.~\ref{speca}.
The spectral evolution differs also qualitatively from that for 
$\theta=0.1\degr$, 
for the blast wave deceleration causes an increase of the Doppler factor
as long as $\mu < \beta$, which in this example holds as long as the 
observed time t$\le 7\cdot 10^6$ sec.}
\label{specc}
\end{figure*}

\begin{figure*}
\resizebox{\hsize}{!}{\includegraphics{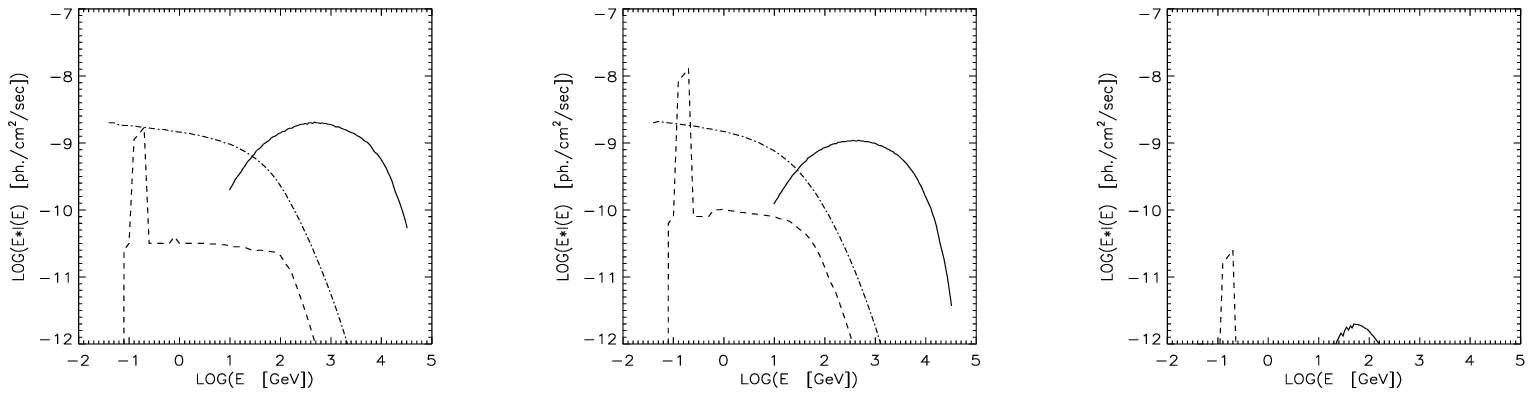}}
\resizebox{\hsize}{!}{\includegraphics{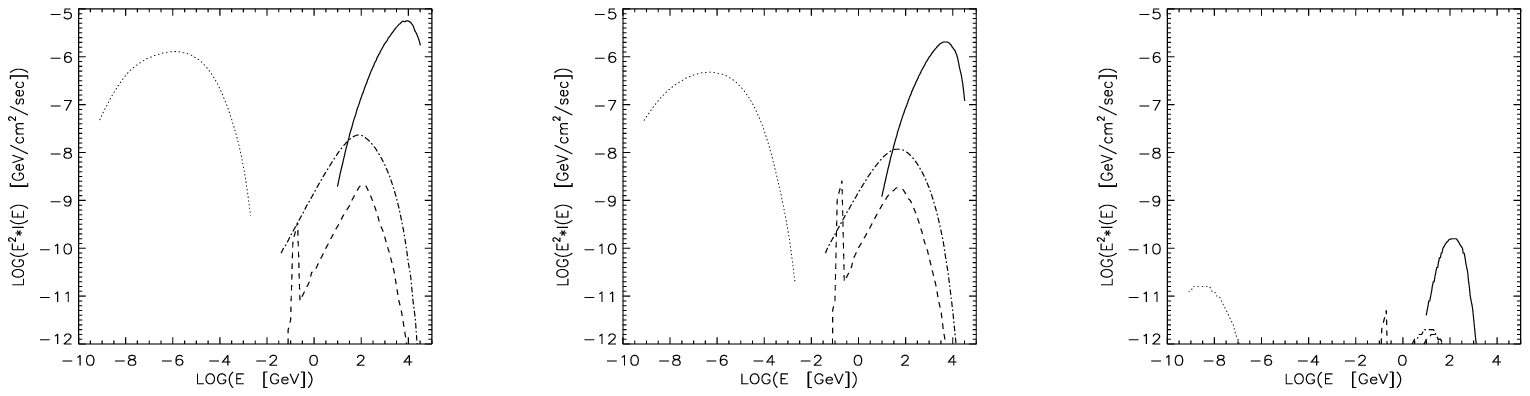}}
\caption{Spectral evolution of a relativistic blast wave having
traversed a gas cloud
of density $n_{\rm i}^\ast$=0.2 cm$^{-3}$ and thickness $5\cdot 10^{16}$ cm. The other
parameters are: $\Gamma_0$=300, d=$10^{14}$ cm, R=$3\cdot 10^{14}$ cm,
B= 2.0 G, $n_{\rm b}$=$10^9$ cm$^{-3}$,
for an AGN at z=0.5 viewed at an angle $\theta=0.1\degr$. The panels show from left to
right the spectra after 0.05 hours, 0.5 hours, and 5 hours
observed time. The Lorentz factor of the blast wave did
virtually not change, hence repeated cloud crossings would produce
multiple outbursts with the same spectral evolution.}
\label{spece}
\end{figure*}
\begin{figure*}
\resizebox{\hsize}{!}{\includegraphics{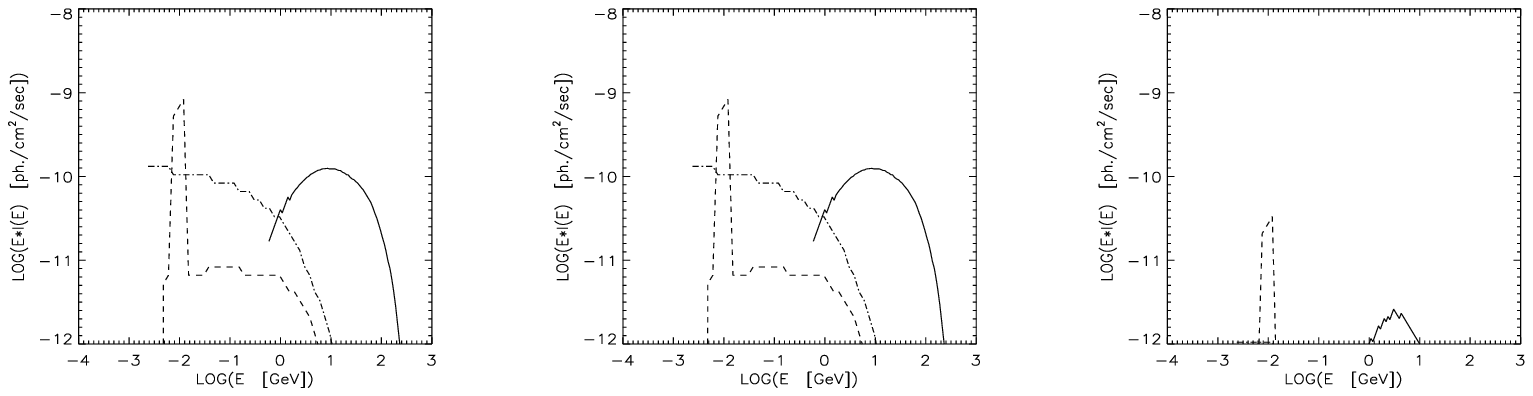}}
\resizebox{\hsize}{!}{\includegraphics{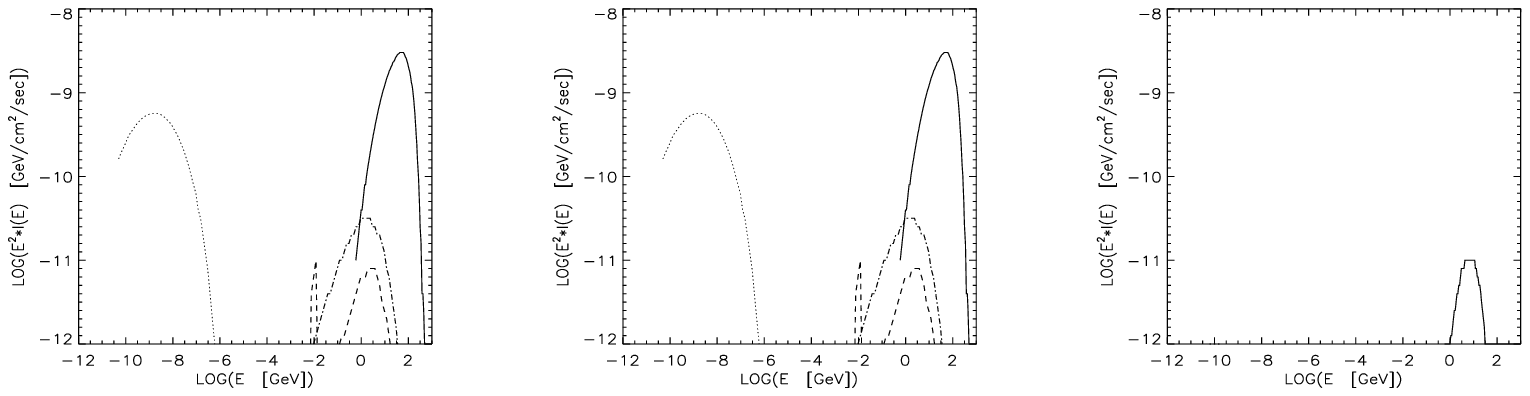}}
\caption{The same system as in Fig.~\ref{spece} viewed by an
observer located 2\degr off the blast wave direction of motion
($\mu$=0.99939).
Note that now the panels show the spectra after 0.5 hours, 5 
hours, and 50 hours of observed time.
Here also d=$3\cdot 10^{14}$ cm, R=$4\cdot 10^{15}$ cm, $n_{\rm i}^\ast$=1 cm$^{-3}$, and
$\Gamma$=30
to make the source appear as bright as before. The choice of the initial
Lorentz factor was done to maximize the observable flux.}
\label{specg}
\end{figure*}

In Fig.~\ref{speca} we show the spectral evolution of high energy emission
from a collimated blast wave for a homogeneous external medium. Even for
very moderate ambient gas densities the high energy emission will be very
intense. Some general characteristics of our model are visible in this figure.
The $\pi^0$-decay component dominates the bolometric luminosity. This
is a direct consequence of the hadronic origin of emission, as the source
power available to the leptonic emission processes is always less than the
pion source power, for the neutrinos carry away part of the energy.

The energy loss time scale of the high energy electrons is always smaller
than that of the protons, and therefore the X-ray synchrotron intensity will
follow dlosely variations in the TeV $\pi^0$-decay emission. Thus there should be
a general relation between the X-ray and the TeV appearance of sources 
with jets pointing towards us. Note, however, that an observer may see an imperfect correlation if a) the synchrotron and the $\pi^0$-decay spectra 
are not observed near their $\nu F_\nu$-peaks, or b) inverse Compton scattering
contributes significantly to the X-ray emission, or c) strong absorption
shifts the apparent $\nu F_\nu$-peak of the $\pi^0$-decay emission.
Correlations between the X-ray and TeV \gr emission of BL Lacertae objects
have indeed been observed in case of Mkn 421 (Buckley et al.
\cite{buc96}) and Mkn 501 (Catanese et al. \cite{cat97}, Aharonian et al.
\cite {aha99}, and Djannati-Ata\"\i\ et al. \cite{dja99}), but at least for
Mkn 421 recent data show that there is no one-to-one correlation during
flares (Catanese et al. \cite{cat99}).

The high Lorentz factors in our model imply that the observed emission
depends more strongly on the aspect angle than in conventional jet models
with $\Gamma \simeq $10. This is shown in Fig.~\ref{specc}, which displays
the spectral evolution of a source viewed at an aspect angle of 2$^\circ$
($\mu$=0.99939).
The model parameters have been changed slightly compared with those
used for the head-on case (Fig.~\ref{speca}), mostly to make the source 
brighter. The fundamental difference to the $\theta\simeq 0\degr$ case is, however, that
the Doppler factor D increases with decreasing $\Gamma$ when $\mu >\beta$.
In contrast to Fig.~\ref{speca}, where always $D>\Gamma$, here D increases
from an initial value of $\sim 5$ to a maximum of $D_{\rm max}
=(1-\mu^2)^{-0.5}
\simeq 29$. As a consequence the apparent evolution of the source spectrum is 
much slower, and the peak energies are smaller and don't change much during
the evolution. The synchrotron peaks are located in the optical, and the
x-ray luminosity is small.

Strong TeV emission implies that the systems need to be observed head-on.
Even for $\Gamma =200$ an aspect angle $\theta \le 1^\circ$ is required.
Viewed at aspect angles of a few degrees the sources look like the typical
hard spectrum EGRET sources: the spectrum peaks somewhere in the GeV region,
the sources reach their maximum luminosity when the blast wave has decelerated
to Lorentz factors around $\sim \sin^{-1} \theta$, the luminosity spectrum
displays a minimum in the keV to MeV range, and the
temporal evolution is slower.

The dependence of the $\nu F_\nu$ peak of the high energy \gr
emission on the aspect angle Eq.~(\ref{pimax}) also implies a general 
relation between the observed $E_{\rm \pi^0,max}$ and the apparent
superluminal velocities
in VLBI images. Using Eq.~(\ref{pimax}) we may write
$$
E_{\rm max,TeV}\, 10^4\simeq {1\over {1-\beta\mu}}$$
\begin{eqnarray}
\Rightarrow\qquad&\mu\simeq \left(1-{1\over {10^4\,E_{\rm max,TeV}}}\right)
\left(1-{1\over {2\,\Gamma_0^2}}\right)^{-1}\nonumber \\
\Rightarrow\qquad&\mu>1-{1\over {10^4\,E_{\rm max,TeV}}}\hfill
\label{mulim}
\end{eqnarray}
From this limit we derive for the apparent velocity
$$\beta_{\rm SL}(t) \simeq {{0.014}\over {E_{\rm max,TeV}}}\left(
{{1\over {10^4\,E_{\rm max,TeV}}}+{1\over {2\,\Gamma(t)^2}}}\right)^{-1}$$
\begin{equation}
\ \simeq 0.028\,{{\Gamma(t)^2}\over {E_{\rm max,TeV}}}
\qquad {\rm for}\ \Gamma(t) \la 70\,\sqrt{E_{\rm max,TeV}}
\end{equation}
This is in accord with the observation that detected EGRET sources, i.e. primarily MeV
-- GeV peaked emitters, often show rapid superluminal motion,
whereas the only TeV peaked \gr source with good VLBI coverage,
namely Mkn 421, displayed subluminal motion in the time range of 1994 to 1997
(Piner et al. \cite{pin99}).

All spectra exhibit a strong annihilation line at an energy of 
$D\cdot 511\,$keV.
The copiously produced positrons cool down to thermal energies before 
annihilating, and a relatively narrow line is produced at a given time.
However, since the Doppler factor may vary during the integration time
of current \gr detectors (typically three weeks (!) in case of COMPTEL),
the line may look like a broad hump in the data.

The spectra shown so far illustrate the effect of the blast wave deceleration
and the thus modulated injection rate. 
To this we oppose in Fig.~\ref{spece} and Fig.~\ref{specg}
the case of the blast wave interacting with an isolated gas cloud.
The light curves of high energy \gr sources often display phases of activity
lasting for months or years, on which rapid fluctuations are superimposed
(e.g. Quinn et al. \cite{qui99}).  
We propose that the secular variability is related to the existence
or non-existence 
of a relativistic blast wave in the sources, and thus to the availability
of free energy in the system. Strong emission on the other hand requires that
the available kinetic energy of the blast wave is converted into
radiation. Depending on the distribution of the interstellar matter
the kinetic energy of the blast wave may be tapped only every now and then. 
The observed fast variability of high-energy $\gamma$-ray sources would
thus be caused by density inhomogeneities in the interstellar medium
of the sources. The situation displayed in Fig.~\ref{spece} would then correspond
to one of the rapid outbursts in the $\gamma$-ray light curves of AGN, which
would generally follow the gas density variations in the volume
traversed by the blast wave. 
  
\begin{figure}
\resizebox{\hsize}{!}{\includegraphics{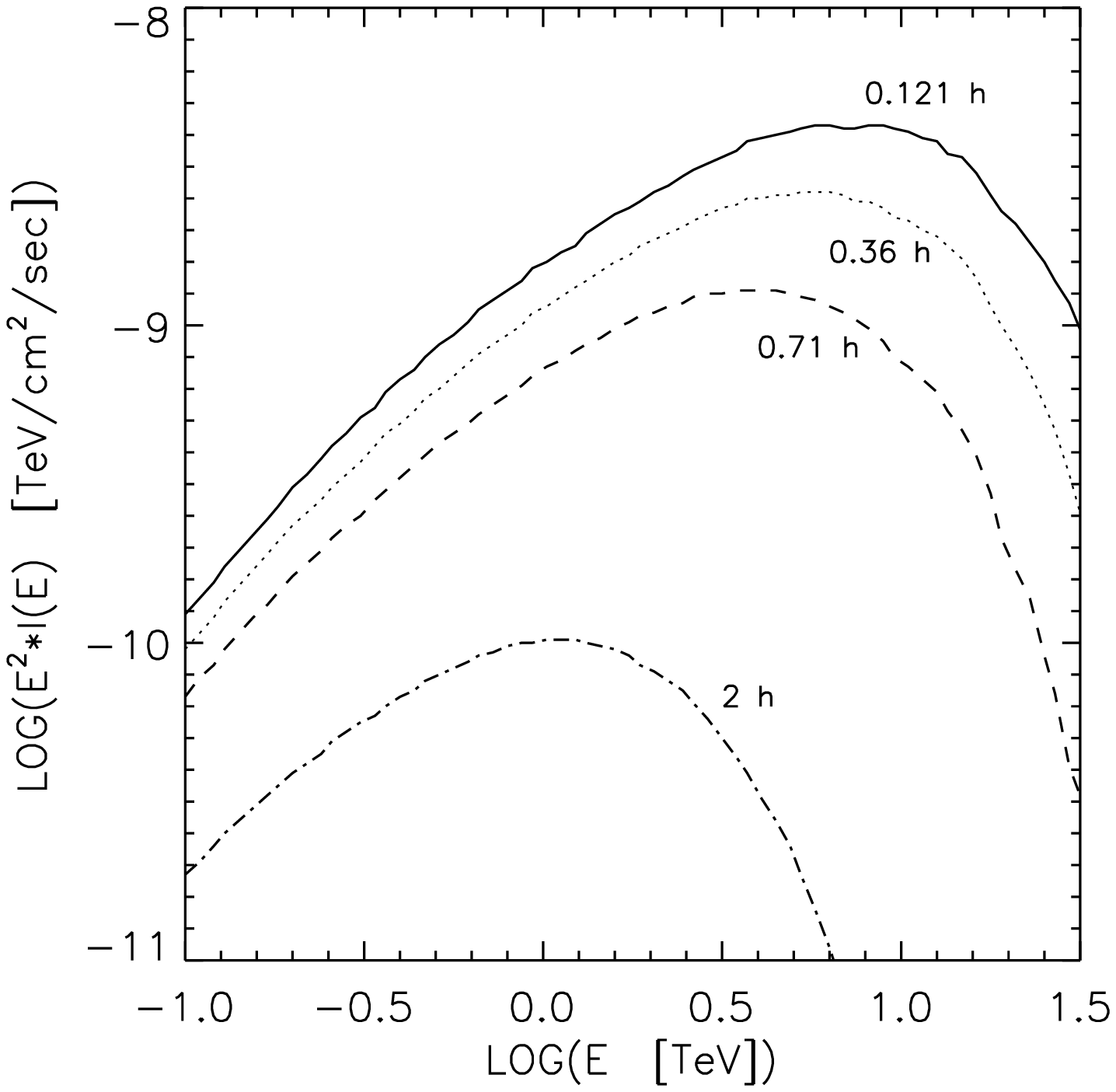}}
\caption{Evolution of the TeV spectra for a single outburst with similar 
parameters as in Fig.\ref{spece}. The curves represent the spectra at different
times after the beginning of the cloud crossing. The labels refer to observed time.}
\label{tevspec}
\end{figure}
\begin{figure}
\resizebox{\hsize}{!}{\includegraphics{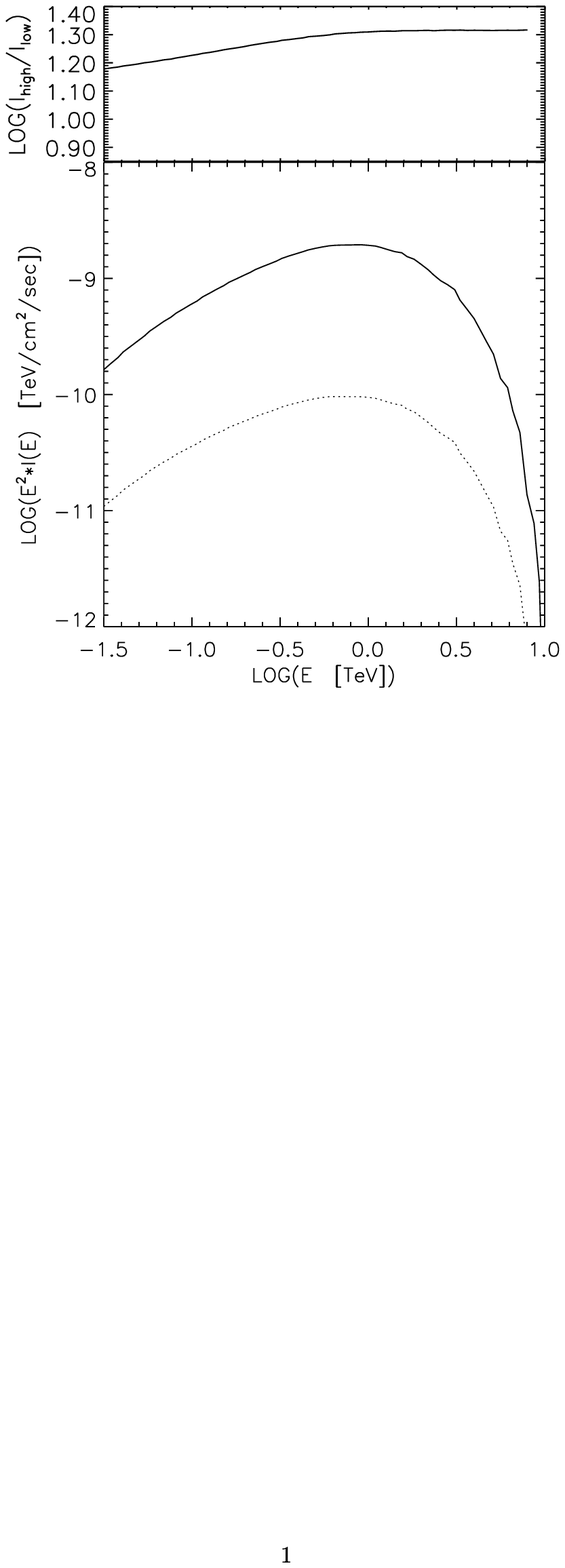}}
\caption{Evolution of the TeV spectra for multiple outbursts
resulting from clouds embedded in the interstellar medium. The interstellar
density $n_{\rm i}^\ast$ was assumed as $10^{-2}\ {\rm cm^{-3}}$ in the
intercloud medium and $0.5\ {\rm cm^{-3}}$ in the clouds, which were 
supposed to be 1 pc in size with a cloud-to-cloud distance of 10 pc.
Here also $\Gamma$=100,
d=$10^{14}\ {\rm cm}$, R=$3\cdot 10^{14}\ {\rm cm}$, $n_{\rm b}=3\cdot 10^9\ 
{\rm cm^{-3}}$, $\theta=0.1\degr$, B=2.0 G, and z=0.5. The modeled data have
been divided into two groups according to the flux
above 0.3 TeV: the flare spectra, here shown as solid line, and the quiet 
spectra, plotted as a dotted line. The top panel show the ratio of the two,
which is compatible with a constant for energies above 0.3 TeV,
thus reproducing the HEGRA result for Mkn 501 (Aharonian et al. \cite{aha99}).}
\label{tevrat}
\end{figure}

The large collection area of currently operating \v Cerenkov telescopes
permits studies of spectral variability on very short timescales.
The observational situation is disturbing, however, since data taken with
different telescopes can be seemingly inconsistent. As an example,
both the CAT team
and the HEGRA group performed observation of Mkn 501 in the time of March 1997
to October 1997. The CAT data show a statistically significant correlation
between the spectral hardness and flux in the energy range between
0.33 TeV and 13 TeV. The HEGRA data on the other hand give no evidence 
in favor of
a correlation in the energy range between 1 TeV and 10 TeV. 

Both behaviours could be reproduced with our model. In Fig.~\ref{tevspec} we
show the spectral variability at TeV energies only. This plot is 
essentially a blow-up of Fig.~\ref{spece}. It is obvious that the rise time
of an outburst is much smaller than the fall time. Also in the declining phase
the source undergoes strong spectral evolution, a consequence of the
proton energy loss timescale being smaller than the escape timescale
in this example. Would the source repeat
the outbursts every, say, five hours, we would definitely observe hard
flare spectra and softer quiet phase spectra.

Were the clouds bigger, so that the cloud crossing time in the blast wave frame
is longer than the proton energy loss timescale, and would we allow for
a dilute intercloud medium, both during the passage through the cloud and 
while traversing the intercloud medium the TeV spectra would approach a steady-state. If we further chose the parameters such that the 
blast wave deceleration is small,
we would reproduce the HEGRA result of variability without spectral evolution
in the TeV emission from Mkn 501. This is shown in Fig.~\ref{tevrat}, 
where the modeled data have been divided into two groups according to the flux
above 0.3 TeV: the flare spectra and the quiet spectra. This is the same
classification that was used in the HEGRA data analysis, except their use
of five groups instead of two. It is obvious that the spectra in both groups
are virtually identical in the energy range that was used for the 
classification. 

\section{Conclusions}
We have shown that a relativistic blast wave can sweep-up ambient
matter via a two-stream instability which provides relativistic
particles in the blast wave without requiring any acceleration process.
We further have demonstrated that the blast waves can have a deceleration
timescale that allows escape from the host galaxy, and hence allows
the formation of giant radio lobes without invoking additional energisation.
While the relativistic blast wave is rather long-lived, the swept-up
relativistic particles in it have short radiation lifetimes. The abundance
of such particle, and hence the intensity of \gr emission, traces 
the density profile of the traversed interstellar gas.
We imagine that the secular variability of the \gr emission of AGN is
related to the existence or non-existence 
of a relativistic blast wave in the sources, and thus to the availability
of free energy in the system.  
The observed fast variability on the other hand would
be caused by density inhomogeneities in the interstellar medium
of the sources. Since we do not consider 
any re-acceleration of particles in the blast wave, the evolution of
particles and the blast wave is completely determined by the initial
conditions. 

The high energy emission of a relativistic blast wave moving approximately
towards the observer has
characteristics typical of BL Lacertae objects. In particular, 

\noindent
-- the high energy spectra are very hard with photon indices $<2$, in
accord with the unspectacular appearance of TeV-bright sources at GeV
energies (Buckley et al. \cite{buc96})

\noindent
-- as can be seen in Fig.~\ref{spece}, observable increase and decrease of 
intensity at TeV energies can be produced on sub-hour time scales,
in accord with the observed
variability time scales of Mkn 421 (Gaidos et al. \cite{gai96})

\noindent
-- for multiple outbursts the intensity can follow the variation of
the ambient gas density with little spectral variation, similar to the observed
behaviour of Mkn 501 (Aharonian et al. \cite{aha99}) 

\noindent
-- x-ray synchrotron emission 
is produced in parallel to the $\gamma$-rays as was observed from Mkn 501
(Pian et al. \cite{pia98}).

\begin{acknowledgements}
Partial support by the Verbundforschung, grant {\it DESY-05AG9PCA}, is gratefully acknowledged.
\end{acknowledgements}

\end{document}